\RequirePackage{fix-cm}
\documentclass[twocolumn,epjc3]{svjour3}  
\smartqed

\usepackage{amsfonts}
\usepackage{amssymb}
\usepackage{amsmath}
\usepackage[titletoc]{appendix}
\usepackage{cite}
\usepackage{cancel}
\usepackage{soul}
\RequirePackage{graphicx}
\usepackage{dcolumn}% Align table columns on decimal point
\usepackage{bm}% bold math
\usepackage{slashed}

\usepackage{IEEEtrantools}
\usepackage{csquotes}
\usepackage{makecell}
\usepackage{color}
\usepackage[version=4]{mhchem}
\usepackage{comment}
\usepackage{multirow}

\usepackage[colorlinks = true,
            linkcolor = blue,
            urlcolor  = blue,
            citecolor = blue,
            anchorcolor = blue]{hyperref}
\usepackage{tikz,xcolor}
            
\DeclareOldFontCommand{\tt}{\normalfont\ttfamily}{\mathtt}
\definecolor{lime}{HTML}{A6CE39}
\DeclareRobustCommand{\orcidicon}{
	\begin{tikzpicture}
	\draw[lime, fill=lime] (0,0) 
	circle [radius=0.16] 
	node[white] {{\fontfamily{qag}\selectfont \tiny ID}};
	\draw[white, fill=white] (-0.0625,0.095) 
	circle [radius=0.007];
	\end{tikzpicture}
	\hspace{-2mm}
}

\newcommand{\DSfs}{\mbox{DS-50}}
\newcommand{\SOne}{\mbox{S1}}
\newcommand{\STwo}{\mbox{S2}}

\foreach \x in {A, ..., Z}{\expandafter\xdef\csname orcid\x\endcsname{\noexpand\href{https://orcid.org/\csname orcidauthor\x\endcsname}
			{\noexpand\orcidicon}}
}

\journalname{Prepared for EPJC}

\newcommand{\APC}{APC, Universit\'e de Paris, CNRS, Astroparticule et Cosmologie, Paris F-75013, France}
\newcommand{\AQLNGS}{INFN Laboratori Nazionali del Gran Sasso, Assergi (AQ) 67100, Italy}
\newcommand{\AQGSSI}{Gran Sasso Science Institute, L'Aquila 67100, Italy}

\newcommand{\AstroCeNT}{AstroCeNT, Nicolaus Copernicus Astronomical Center, 00-614 Warsaw, Poland}
\newcommand{\Augustana}{Physics Department, Augustana University, Sioux Falls, SD 57197, USA}
\newcommand{\Belgorod}{Radiation Physics Laboratory, Belgorod National Research University, Belgorod 308007, Russia}
\newcommand{\BHSU}{School of Natural Sciences, Black Hills State University, Spearfish, SD 57799, USA}

\newcommand{\CAUniPHY}{Physics Department, Universit\`a degli Studi di Cagliari, Cagliari 09042, Italy}
\newcommand{\CAINFN}{INFN Cagliari, Cagliari 09042, Italy}

\newcommand{\CPPM}{Centre de Physique des Particules de Marseille, Aix Marseille Univ, CNRS/IN2P3, CPPM, Marseille, France}
\newcommand{\CTLNS}{INFN Laboratori Nazionali del Sud, Catania 95123, Italy}
\newcommand{\ENUniCEE}{Engineering and Architecture Faculty, Universit\`a di Enna Kore, Enna 94100, Italy}

\newcommand{\FNAL}{Fermi National Accelerator Laboratory, Batavia, IL 60510, USA}

\newcommand{\GEUni}{Physics Department, Universit\`a degli Studi di Genova, Genova 16146, Italy}
\newcommand{\GEINFN}{INFN Genova, Genova 16146, Italy}

\newcommand{\Hawaii}{Department of Physics and Astronomy, University of Hawai'i, Honolulu, HI 96822, USA}
\newcommand{\Houston}{Department of Physics, University of Houston, Houston, TX 77204, USA}
\newcommand{\IHEP}{Institute of High Energy Physics, Beijing 100049, China}

\newcommand{\JINR}{Joint Institute for Nuclear Research, Dubna 141980, Russia}
\newcommand{\Krakow}{M. Smoluchowski Institute of Physics, Jagiellonian University, 30-348 Krakow, Poland}
\newcommand{\Kurchatov}{National Research Centre Kurchatov Institute, Moscow 123182, Russia}

\newcommand{\LNFINFN}{INFN Laboratori Nazionali di Frascati, Frascati 00044, Italy}
\newcommand{\LNHB}{Universit\'e Paris-Saclay, CEA, List, Laboratoire National Henri Becquerel (LNE-LNHB), F-91120 Palaiseau, France}

\newcommand{\LPNHE}{LPNHE, CNRS/IN2P3, Sorbonne Universit\'e, Universit\'e Paris Diderot, Paris 75252, France}
\newcommand{\Manchester}{The University of Manchester, Manchester M13 9PL, United Kingdom}
\newcommand{\MEPhI}{National Research Nuclear University MEPhI, Moscow 115409, Russia}

\newcommand{\MIINFN}{INFN Milano, Milano 20133, Italy}

\newcommand{\MIUni}{Physics Department, Universit\`a degli Studi di Milano, Milano 20133, Italy}
\newcommand{\MSU}{Skobeltsyn Institute of Nuclear Physics, Lomonosov Moscow State University, Moscow 119234, Russia}
\newcommand{\NAINFN}{INFN Napoli, Napoli 80126, Italy}
\newcommand{\NAUniPHY}{Physics Department, Universit\`a degli Studi ``Federico II'' di Napoli, Napoli 80126, Italy}

\newcommand{\Petersburg}{Saint Petersburg Nuclear Physics Institute, Gatchina 188350, Russia}
\newcommand{\PGUniCBB}{Chemistry, Biology and Biotechnology Department, Universit\`a degli Studi di Perugia, Perugia 06123, Italy}
\newcommand{\PGINFN}{INFN Perugia, Perugia 06123, Italy}
\newcommand{\PIINFN}{INFN Pisa, Pisa 56127, Italy}
\newcommand{\PIUniPHY}{Physics Department, Universit\`a degli Studi di Pisa, Pisa 56127, Italy}
\newcommand{\PNNL}{Pacific Northwest National Laboratory, Richland, WA 99352, USA}
\newcommand{\Princeton}{Physics Department, Princeton University, Princeton, NJ 08544, USA}

\newcommand{\RHUL}{Department of Physics, Royal Holloway University of London, Egham TW20 0EX, UK}
\newcommand{\RMTreINFN}{INFN Roma Tre, Roma 00146, Italy}
\newcommand{\RMTreUni}{Mathematics and Physics Department, Universit\`a degli Studi Roma Tre, Roma 00146, Italy}
\newcommand{\RMUnoINFN}{INFN Sezione di Roma, Roma 00185, Italy}
\newcommand{\RMUnoUni}{Physics Department, Sapienza Universit\`a di Roma, Roma 00185, Italy}

\newcommand{\SSUniCHP}{Chemistry and Pharmacy Department, Universit\`a degli Studi di Sassari, Sassari 07100, Italy}

\newcommand{\UCDavis}{Department of Physics, University of California, Davis, CA 95616, USA}
\newcommand{\UCLA}{Physics and Astronomy Department, University of California, Los Angeles, CA 90095, USA}
\newcommand{\UCRiverside}{Department of Physics and Astronomy, University of California, Riverside, CA 92507, USA}
\newcommand{\UMass}{Amherst Center for Fundamental Interactions and Physics Department, University of Massachusetts, Amherst, MA 01003, USA}

\newcommand{\USP}{Instituto de F\'isica, Universidade de S\~ao Paulo, S\~ao Paulo 05508-090, Brazil}
\newcommand{\VTech}{Virginia Tech, Blacksburg, VA 24061, USA}
\newcommand{\kings}{Physics, Kings College London, Strand, London WC2R 2LS, UK}

\usepackage{enumitem}
\newlist{enumcompactenum}{enumerate}{3}
\setlist[enumcompactenum]{topsep=0pt,partopsep=0pt,itemsep=0pt,parsep=0pt}
\setlist[enumcompactenum,1]{label=\arabic*}
\setlist[enumcompactenum,2]{label=\alph*}
\setlist[enumcompactenum,3]{label=\roman*}

\makeatletter
\renewcommand{\thanksref}[1]{\nolinebreak\textsuperscript{\ref{#1}}\nolinebreak\checknextarg}
\newcommand{\checknextarg}{\@ifnextchar\bgroup{\nolinebreak\gobblenextarg}{}}
\newcommand{\gobblenextarg}[1]{ \textsuperscript{\nolinebreak\hspace{-4pt}\mbox{\nolinebreak$^,$\nolinebreak\ref{#1}\nolinebreak}\nolinebreak} \@ifnextchar\bgroup{\gobblenextarg}{}}

\begin{document}

\title{Search for low mass dark matter in DarkSide-50: the bayesian network approach
}

\author{The DarkSide-50 Collaboration$^\text{\normalfont a,1}$}

\thankstext{e1}{e-mail: ds-ed@lngs.infn.it}

\institute{See back for author list \label{addr1}}

\date{2 Feb 2023}

\maketitle

\begin{abstract}

We present a novel approach for the search of dark matter in the DarkSide-50 experiment, relying on Bayesian Networks. This method incorporates the detector response model into the likelihood function, explicitly maintaining the connection with the quantity of interest. No assumptions about the linearity of the problem or the shape of the probability distribution functions are required, and there is no need to morph signal and background spectra as a function of nuisance parameters.
By expressing the problem in terms of Bayesian Networks, we have developed an inference algorithm based on a Markov Chain Monte Carlo to calculate the posterior probability. A clever description of the detector response model in terms of parametric matrices allows us to study the impact of systematic variations of any parameter on the final results.
Our approach not only provides the desired information on the parameter of interest, but also potential constraints on the response model. Our results are consistent with recent published analyses and  further refine the parameters of the detector response model.


\end{abstract}

\tableofcontents

\section{Introduction}
\label{sec:intro}

The frontiers of physics are often explored conducting experiments at the limit of the detector sensitivity. In such a circumstance it is extremely important to control and correctly evaluate the effect of systematic uncertainties. This is particularly relevant when hints of a new signal are sought in event samples contaminated by a significant background that might amplify the impact of systematic effects and thus dilute the relevance of the observation.

The search for dark matter candidates with direct detection experiments is one of such cases where a very rare and feeble signal is searched for in a huge background of events, mostly induced by cosmic-rays and natural radioactivity. In this scenario, robust results in terms of projected experimental sensitivities, as well as exclusion limits, or hints of new physics are only achieved after a deep understanding of all relevant experimental effects.   

In the past decade, a number of experiments have been searching for dark matter with masses in the range of a few GeV/c$^2$ and below, pushing the sensitivity to their experimental limit in terms of total exposure and energy threshold. The current status of the art is led by experiments exploting scintillation light and ionization charge~\cite{Agnes:2018ves,Aprile:2019jmx,Aprile:2019xxb,PandaX-4T:2021bab,DarkSide-50:2022qzh,DarkSide:2022dhx} or heat~\cite{CRESST:2019jnq,CDEX:2021cll}. In particular, the DarkSide-50 (\DSfs) experiment~\cite{DarkSide-50:2022qzh,DarkSide:2022dhx}, a large volume dual phase argon Time Projection Chamber, has set the strongest constraint on the spin-independent cross section for dark matter masses below $3.6$ GeV/c$^2$.

In the search for very rare phenomena, the optimal strategy is to identify and reject all background events while retaining a significant fraction of signal events.
In this ideal case, often referred to as zero-background search, a handful of candidate events is enough to claim a discovery. 
More commonly, it is not possible to efficiently separate signal from background events and a different strategy is needed. A refined data analysis has to be put in place to compute the probability that the observed candidate events are due to new physics. Such analyses are based on the so called likelihood function that provides a parametric model for the probability to make an observation assuming certain hypotheses. A detailed and sound description of detector effects and background sources has to go in the likelihood before possible discrepancies observed in the data can be ascribed to new phenomena. 

The likelihood function typically depends on a number of parameters, some of which, called parameters of interest, represent the physics quantities relevant for the measurements, some others, called nuisance parameters, encode the detector response and the background properties.
Although the likelihood function is at the basis of the inferential processes both in the Bayesian and Frequentist approaches, the impact of the nuisance parameters on the parameters of interest is treated very differently.

In the Frequentist case, the likelihood function is maximised and an approximated dependence of the parameter of interest on the nuisance parameters is obtained through the so called profiling procedure\footnote{ This procedure represents the most common approach in Frequentistic analyses to take systematic uncertainties into account.} \cite{Cowan:2010js}. However, the dependence of the parameter of interest on these parameters is in general non-linear, and very complicated. For these reasons, analysis approaches based on the profiling of the likelihood of the experiment, valid under linearity, symmetry, or ``Gaussianity'' assumptions~\cite{ParticleDataGroup:2020ssz}, might not be able to reproduce in an accurate way the propagated uncertainties on the parameter of interest.  

In the Bayesian approach, a joint posterior probability density function (\emph{pdf}) for all parameters is constructed from the likelihood function and the prior \emph{pdf}. The impact of systematic uncertainties on the parameters of interest is obtained by integrating (marginalising) the posterior \emph{pdf} over the nuisance parameters. This procedure is straightforward and well motivated by probability theory. The main criticism to the Bayesian method related to the intrinsic `subjective' nature of these prior \emph{pdf}s is not relevant in this case, since the \emph{pdf}s for the nuisance parameters are normally determined by ancillary measurements.  

Independently of the inferential approach, the construction of the likelihood function tailored on the specificity of the measurement is a challenging task.
Given the complexity of the detector response and the diverse origin of the background sources, the likelihood function is often sampled with Monte Carlo (MC) simulations. 
Expected differential event rates are constructed as a function of some relevant observables in the form of histograms of events. These histograms, known as templates, are an approximation of the \emph{pdf}s for the observables given some specific set of parameters describing the experiment. Different configurations are explored by computing new templates from MC simulations run with different sets of parameters. Given the typical size of millions of events needed to produce accurate templates, this procedure is very time-consuming. To overcome this limitation, a few points in the parameter space are chosen as representative of the systematic variations with respect to the best model. New templates are derived from the computed ones by linear interpolation or more advanced morphing techniques \cite{Baak:2014fta}. 

This work shows that, under certain circumstances, the likelihood function can be expressed as an analytical or semi-analytical function of the relevant parameters, making it unnecessary to run large MC simulations to account for systematic effects. A similar approach, not cast in the Bayesian inference language, has been presented in Ref.~\cite{Aalbers:2020iej,LUX:2022vee,James:2022sgg}. Alternative analyses incorporating machine learning techniques have been showcased in Ref.~\cite{Coarasa:2022zak, Herrero-Garcia:2021goa}.  In addition, our paper describes a new technique based on probabilistic graphical models, also known as Bayesian Networks (BN)~\cite{pearlbook, jensenbook, DAgostini:2003syq, kollerbook}, to take into account systematic uncertainties. 
In a BN the probabilistic relations between the physical quantities and the observations are evident, and the dependence on detector effects and background contributions is made explicit, allowing for reasoning about causes and effects within the model~\cite{ref31, ref32, ref33, ref34, ref35, ref36, Peters:2022afo}.
To our knowledge, this approach has first been proposed for parametric inference in presence of data points affected by systematic uncertainties in Ref.~\cite{DAgostini:2005mth}, and later used to describe the Bayesian unfolding in Ref.~\cite{2010arXiv1010.0632D}.

This technique has the advantage over the widely used profiling methods to be exact in terms of uncertainty propagation~\cite{DAgostini:2004kis}, to not rely on template morphing, and to properly take into account cross correlations between parameters and phase space regions.
In addition, if the physical parameters describing the detector response model and constrained by calibrations are retained as parameters inside the likelihood function, this method gives the possibility of verifying the goodness of the calibrations and, \emph{a posteriori}, further constrains the detector response model.

The rest of this article proceeds as follow. In Sec.~\ref{sec:concept} we describe the BN method and its implementation for an oversimplified DM direct detection experiment.
The method is applied to the low mass DM analysis of the DarkSide-50 (\DSfs) experiment. The description of the experimental setup and the data-set employed in this study is contained in Sec.~\ref{sec:DS50}.

Section~\ref{sec:detector-effects_responsemodel} describes the detector response model and the implementation via the BN method. In Sec.~\ref{sec:likelihood} we show the \DSfs\ likelihood, while in Sec.~\ref{sec:fitproc} we describe the technical implementation.
Finally, in Sec.~\ref{sec:results}, we illustrate the results of this method in terms of sensitivity and exclusion limits for the low mass DM analysis, exploiting both the nuclear recoil (NR) and Migdal effect (ME)~\cite{Migdal:1941} signals. The ME signal has been recently studied in Ref.~ \cite{Bernabei:2007jz,Ibe:2017yqa,Dolan:2017xbu,Bell:2019egg,Baxter:2019pnz,Essig:2019xkx,Liang:2019nnx,GrillidiCortona:2020owp,Liu:2020pat,Dey:2020sai,Knapen:2020aky,Bell:2021zkr,Acevedo:2021kly,Wang:2021oha}, and exploited in the search for light DM in Ref.~\cite{LUX:2018akb, EDELWEISS:2019vjv,CDEX:2019hzn,XENON:2019zpr,COSINE-100:2021poy,CDEX:2021cll,DarkSide:2022dhx}.
Being these signals a combination of NR and electronic recoil (ER) energy releases, we also test the simultaneously handling of the detector calibrations of the two channels.
We obtain comparable sensitivity with respect to the recent published analysis~\cite{DarkSide:2022dhx}. We also show that making the likelihood explicitly dependent on the detector parameters gives us a much better control over the systematic effects and allows improving the knowledge on the detector response parameters exploiting the new data. 
In Sec.~\ref{sec:conclusions} we draw our conclusions.

\section{Concept and method}
\label{sec:concept}
\subsection{Graphical method for Bayesian inference}

According to the Bayesian Networks method, the probability density function of a collection of random variables can be represented by a network made of arrows and nodes where:
\begin{enumerate}
    \item the nodes represent the variables;
    \item a solid arrow between two nodes represents a probabilistic link between the two variables;
    \item a dashed arrow between two nodes represents a deterministic link between the two variables;
    \item a gray node indicates the corresponding variable has been observed. 
\end{enumerate}

\begin{figure}
    \centering
    \includegraphics[width = 0.35\textwidth]{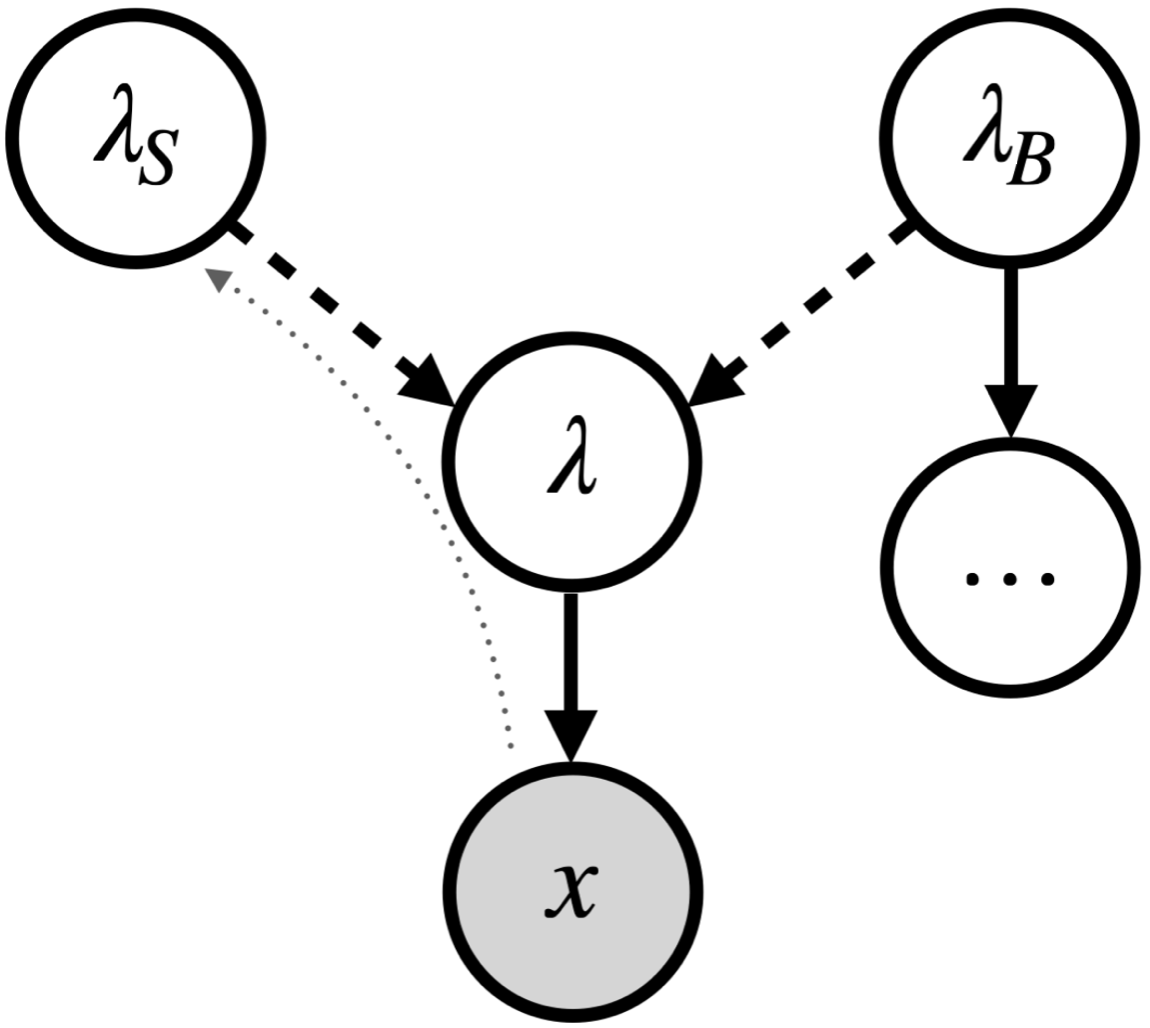}
    \caption{\label{fig:BN_poisson}Bayesian network for a Poisson process with intensity $\lambda$ and observed number of events $x$. The intensity $\lambda$ is in turn the sum of a signal contribution $\lambda_S$ and a background contribution $\lambda_B$, and thus the links connecting $\lambda_S$ and $\lambda_B$ to $\lambda$ are deterministic. The dots denote a possible additional portion of the network representing previous measurements, e.g. during calibration, of $\lambda_B$. The inferential process is an information flow (gray dotted arrow) from the observed node $x$, denoted in gray, to the parameter of interest $\lambda_S$.}
\end{figure}

To briefly illustrate the method, let us take as an example the search for a rare process in presence of a background contribution~\cite{Astone:1999wp,DAgostini:2015nao}. This is typically described by a Poisson likelihood, defined as
\begin{equation}
    p(x|\lambda = \lambda_B + \lambda_S) =
    \frac{(\lambda_B + \lambda_S)^x}{x!} e^{-(\lambda_B + \lambda_S)},
    \label{eq:poisson}
\end{equation}
where $x$ is the observed datum, $\lambda$ is the intensity of the Poisson process, $\lambda_B$ regulates the intensity of the background contribution, and $\lambda_S$ represents the signal strength parameter we want to determine.
Figure~\ref{fig:BN_poisson} shows the BN for this simple measurement. 
We are interested in inferring $p(\lambda_S|x)$, namely the \emph{pdf} of $\lambda_S$, conditioned to the observation of the node $x$. In other words, looking at the BN of Fig.~\ref{fig:BN_poisson}, we want to determine how the information obtained with the observation of the node $x$ propagates backwards in the network to the parameter of interest $\lambda_S$. In this sense, the Bayesian inferential process can be regarded as an information flow from the observed nodes to the parameters of interest. Once we draw the network, we determine the relations represented by each of the links, and we fix the observed nodes, the rules of probability allow determining the \emph{pdf} of the unobserved nodes conditioned to the measured data $p(\lambda_S, \lambda_B, \lambda| x)$, also called the \emph{posterior} \emph{pdf}, or simply the posterior.
If we are only interested in $\lambda_S$, we can obtain the posterior $p(\lambda_S|x)$ integrating over the nuisance parameters' space.

\subsection{Implementation for dark matter direct detection experiments}

For this work, we consider the dual phase noble liquid TPC approach to search for DM exploiting the ionization channel only~\cite{DarkSide:2014llq, XENON:2018voc, LUX:2016ggv, PandaX:2014mem}. Simplifying as much as possible, the experiment consists in counting the number of detected events in the active volume as a function of the measured amount of detectable quanta produced during the primary event. The number of detectable quanta is in turn a measurement of the energy of the recoiling target particle.
Finally, a histogram of the observed spectrum is obtained.
In order to put constraints on the possible DM signal contribution to the observed spectrum, two main ingredients are needed:
\begin{enumerate}
\item a detector response model describing how the kinetic energy transferred to the target particles is translated into the number of detectable quanta measured by the experimental apparatus. According to the type of experiment, the target material, and the readout system, this response model can also be highly non-linear, and the parameters regulating it are usually constrained by a set of preliminary calibration measurements.
\item a background model describing the relevant background contributing to the final observed spectrum.
\end{enumerate}
These two models allow computing the expected spectrum
starting from the theoretical background and signal spectra.

The common way to implement these two models for the analysis is via a toy MC approach, in which the final expected spectrum and its $\pm\sigma$ (standard deviation) variations are computed before the fit. During the fit, a morphing procedure is used to take into account possible variations of the spectral shape. According to this approach, a new set of nuisance parameters is used; the number of these newly added parameters and their correlations are chosen on the basis of ad hoc and case dependent prescriptions, while their physical interpretations in terms of physical systematic parameters is often lost. In other words, due to the re-parameterization of the likelihood, the post fit results can not easily be interpreted in terms of the original physical parameters describing the systematic effects.

We developed a method to compute the final expected spectrum as an analytical function of the theoretical spectrum and the experimental parameters. Indeed, the probability to measure a certain number $N_q$ of detectable quanta given a certain kinetic recoil energy $E$ and a certain set of experimental parameters $\boldsymbol{\theta}$, can be decomposed as
\begin{align}
    & p(N_q = i | E = E_k, \boldsymbol{\theta}) =  \nonumber \\
    & \sum_j
    p(N_q = i | N_q^{(0)} = j, \boldsymbol{\theta})
    \:
    p(N_q^{(0)} = j | E = E_k, \boldsymbol{\theta}),
\end{align}
where $N_q^{(0)}$ is the originally \emph{produced} number of detectable quanta, namely the number of detectable quanta before applying other detector related effects such as response non linearity, efficiency, or resolution.
Since we have the possibility of computing these two \emph{pdfs} analytically and since the spectra we are dealing with can be treated as vectors, we can express the final expected spectrum $S_i^{fin}$ as the following linear combination of the theoretical spectrum $S_k^{th}$:
\begin{equation}
     S^{fin}_i (N_q = i, \boldsymbol{\theta}) =
     \sum_{j,k} \mathcal{M}^2_{ij} (\boldsymbol{\theta})
    \:
    \mathcal{M}^1_{jk}(\boldsymbol{\theta})
    \:
    S^{th}_k(E = E_k)
    \label{eq:th_to_fin}
\end{equation}
where
\begin{equation}
\begin{cases}
    \mathcal{M}^1_{jk}(\boldsymbol{\theta}) &\equiv
    p(N_q^{(0)} = j | E = E_k, \boldsymbol{\theta}) \\
    \mathcal{M}^2_{ij} (\boldsymbol{\theta}) &\equiv
    p(N_q = i | N_q^{(0)} = j, \boldsymbol{\theta}). \\
\end{cases}
\label{eq:SM_def}
\end{equation}
These two matrices could depend on the background source or on the readout channel.
The ability to compute the entries of $\mathcal{M}^1$ and $\mathcal{M}^2$, called the \emph{smearing matrices}, allows us to obtain the final expected spectrum via Eq.~\eqref{eq:th_to_fin} and to perform the fit on the observed one by means of linear algebra operations. 

\begin{figure*}
    \centering
    \includegraphics[width = 0.40\textwidth]{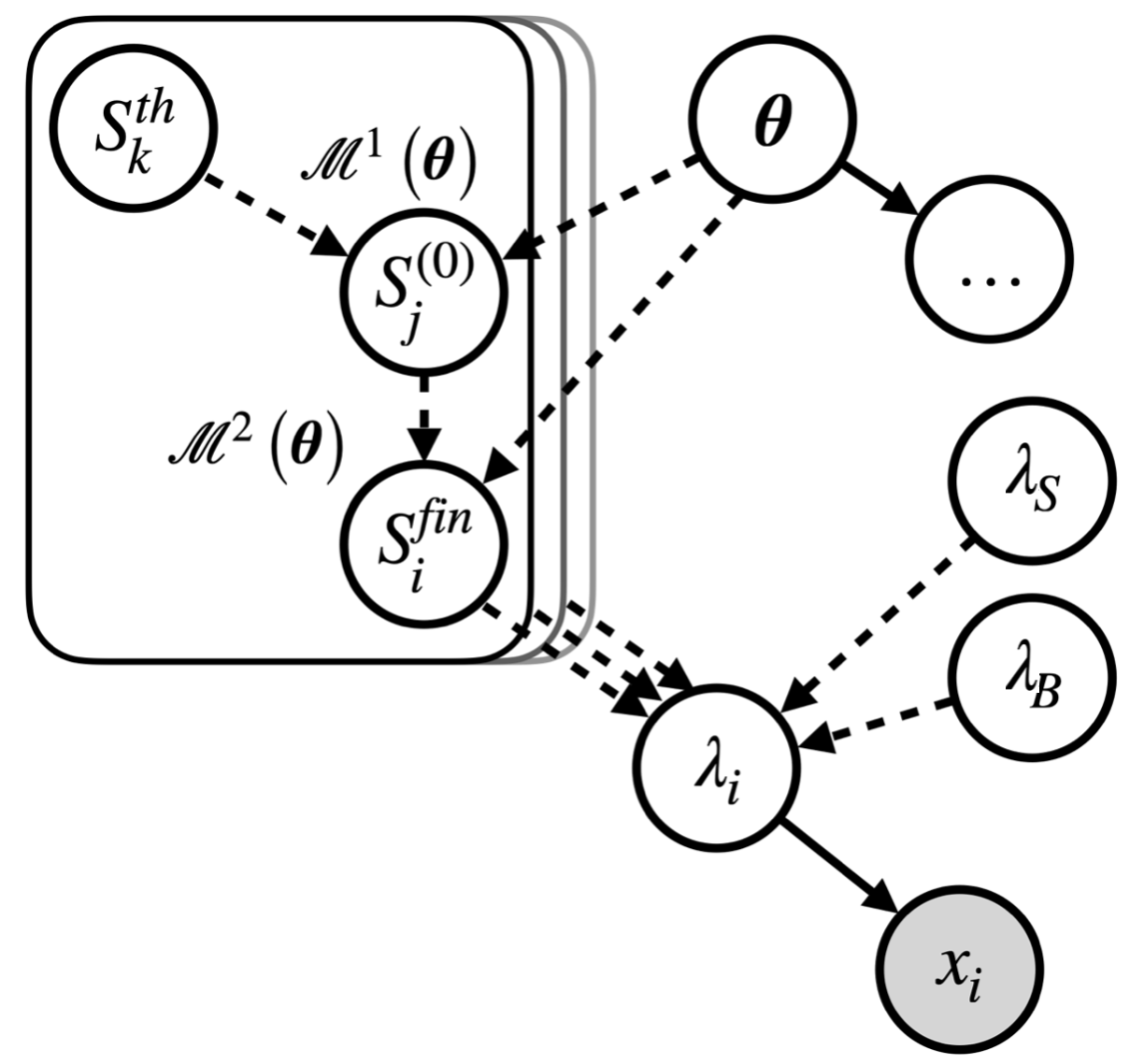}
   \includegraphics[width = 0.48\textwidth]{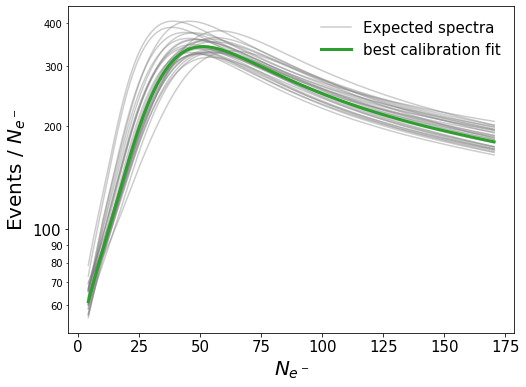}
    \caption{{\bf Left:} Bayesian network for a prototypical dual phase TPC DM direct detection experiment. The node $x_i$ is the observed number of events in the $i$-th bin; $\lambda_i$ is the intensity of the Poisson process in the $i$-th bin; $\lambda_S$ and $\lambda_B$ regulate the strength of the signal and the background contributions, respectively; the boxes in different gray intensities represent the different contributions from all the relevant background or signal sources; $\boldsymbol{\theta}$ are the parameters describing the detector response model; the dots indicate possible preliminary calibration measurements of $\boldsymbol{\theta}$; $\mathcal{M}^1$ and $\mathcal{M}^2$ are defined in Eq.~\eqref{eq:SM_def}.
    {\bf Right:} a possible background expected spectrum ($S_i^{fin}$) for different configurations of the $\boldsymbol{\theta}$ parameters (gray lines). The green curve corresponds to the best calibration fit values of $\boldsymbol{\theta}$.
    }
    \label{fig:BN_tot}
\end{figure*}

Figure~\ref{fig:BN_tot}~(top) shows the Bayesian network describing all the process. In particular, $S_j^{(0)}$ is the intermediate spectrum after applying only the $\mathcal{M}^1$ matrix; the dots indicate possible preliminary calibration measurements constraining $\boldsymbol{\theta}$; the boxes in different gray intensities represent the fact that there are many copies of that portion of the network, one for each background or signal component; $x_i$ indicates the number of observed events in the $i$-th bin, while $\lambda_i$ is the Poisson expected value in the $i$-th bin; $\lambda_S$ and $\lambda_B$ are normalization parameters regulating the intensity of the signal and the background contributions, respectively.
Once the BN is constructed, one can use it in a twofold way. Top-down, going from the parameters of the model to the observation, it can be used to simulate events by sampling for example with a Markov Chain Monte Carlo (MCMC); or bottom-up, going from the observation to the parameters of the model, it can be used to perform the inference and measure the parameter of interest. As an example of this method, Fig.~\ref{fig:BN_tot}~(bottom) shows graphically an intermediate node of the BN during the sampling phase, namely a possible background expected spectrum ($S_i^{fin}$) for different configurations of the $\boldsymbol{\theta}$ parameters (gray lines). In addition, the green curve corresponds to the best calibration values of $\boldsymbol{\theta}$ obtained a \emph{posteriori} after a fit to a background only observation. From this figure, it is clear that each set of calibration parameters produces a background template that cannot be parametrized with simply a normalization and shape variation without losing the correlation among bins.

\section{The DarkSide-50 experiment}
\label{sec:DS50}
The \DSfs\ experiment exploits a dual-phase liquid argon (LAr) time projection chamber (TPC), operated in Italy at the INFN's Laboratori Nazionali del Gran Sasso (LNGS). Two measurable interactions can be observed in the 46.4$\pm$0.7 kg active target: light from scintillation in the liquid (\SOne) and ionization electrons. A 200 V/cm electric field in the LAr volume drifts the ionization electrons up to the gas pocket, where they are extracted by an electric field of 2.8 kV/cm into the gas phase, producing a secondary pulse of light (\STwo) by electroluminescence. The ultraviolet photons from the \SOne\ and \STwo\ signals are converted by a tetraphenyl butadiene wavelength shifter that covers the internal surface of the TPC into visible radiation. Two arrays of 19 3-in photomultipliers (PMTs), located above the anode and below the cadhode, detect these photons. The TPC is enclosed in a stainless steel cryostat and lies inside a 30 t boron-loaded liquid scintillator veto equipped with 110 8-in PMTs, in order to actively reject neutrons. This is surrounded by a 1 kt ultra-pure water Cherenkov veto with 80 PMTs that acts as a cosmic muon veto and as a passive shield against external backgrounds. Further details on the detector can be found in Ref.~\cite{DarkSide:2014llq,DarkSide:2015cqb,DarkSide:2018bpj,DarkSide:2018kuk,DarkSide:2018ppu}.
\subsection{Event selection}
This analysis uses the full \DSfs\ exposure of $\mathcal{E}_{\rm DS} = 653.1$ live-days, from December 12, 2015 to February 24, 2018. 
The event selection is exactly the same as the one used for the recent \DSfs\ publications on the search for low-mass DM \cite{DarkSide:2022dhx, DarkSide-50:2022qzh} and it is described in details in Ref.~\cite{DarkSide-50:2022qzh}.
We define the region of interest (ROI) as where the ionization response is calibrated~\cite{DarkSide:2021bnz} and backgrounds are well-understood~\cite{DarkSide-50:2022qzh}. This corresponds to the NR(ER) energy range [0.06,\,21] keV$_{\textrm{er}}$ ([0.64,\,393] keV$_{\textrm{nr}}$), matching to the number of electron range $N_e$ within [4,\,170]. Below $4e^-$, a non-negligible contribution to the background model of spurious electrons captured by impurities along and re-emitted with a delay is present. 
This analysis considers only single scatter events with a single \STwo\ pulse, except for `echoes' (pulses induced by one or more $e^-$ extracted from the cathode from \SOne\ or \STwo\ photons via photoelectric effect).
Several quality cuts based on \STwo\ $/$ \SOne\ and on the topological distribution and the time profile of the \STwo\ signal are implemented. These cuts reject events with overlapping pulses \cite{DarkSide-50:2022qzh}. Additional cuts are applied in order to remove events with an anomalous start time, associated to random coincidences between very low \SOne\ and \STwo\ pulses from the anode, alpha particles emitted near the TPC walls, whose \SOne\ photons induce \STwo\ pulses by extracting electrons from the cathode, and spurious \STwo\ pulses, induced by electrons captured by impurities. After the quality and selection cuts, the final data-set results in an exposure of $12202\pm180$ kg d.

\subsection{Background model}
\label{sec:bkgd_model}
The ROI is dominated by internal backgrounds, such as the $^{85}$Kr and $^{39}$Ar decays occurring in the LAr fiducial volume, and external backgrounds, including radiation from contaminants in the PMTs and stainless-steel cryostat. Additional negligible backgrounds originate from radiogenic and cosmogenic neutrons, and coherent elastic neutrino-nucleus scattering from solar and atmospheric neutrinos. 
The spectral shapes of the internal backgrounds take into account recent calculations of atomic exchange and screening effects, validated on measured $\ce{^{63}Ni}$ and $\ce{^{241}Pu}$ spectra down to $200$ eV~\cite{PhysRevA.90.012501,Haselschwardt:2020iey}. From 0 eV to $200$ eV a linear uncertainty on such corrections ranging from $25\%$ to $0\%$ is assumed.
Other systematics on the spectral shape come from the ionization response, modeled via Monte Carlo simulations~\cite{DarkSide:2021bnz}, and subdominantly from the uncertainty on the Q-value~\cite{Wang:2021xhn}. 
The external background model is based on simulations of each material component as measured during the material screening campaign. The input for the simulation are given by these measurements and the associated uncertainties, after the correction because of the decrease activity due to the elapsed time at the dataset date. A more detailed description on the background models can be found in Ref.~\cite{DarkSide-50:2022qzh}.

\subsection{DarkSide-50 response model}
The detector calibration and its response has been performed in Ref.~\cite{DarkSide:2021bnz} for both ER and NR energy deposits. 

The ionization yield of an ER or a NR, denoted by the symbol $Q_y$, is defined as the average number of ion-electron pairs surviving recombination per unit of energy. The two models describing the ionization yields, validated and constrained by the calibration measurements \cite{DarkSide:2021bnz}, are a modified version of the Thomas-Imel box model for the ER~\cite{Thomas:1987zz} and the Bezrukov model for the NR~\cite{Bezrukov:2010qa}. The ER ionization yield is given by
\begin{equation}
    Q_y^{ER} = \left[a_1 + a_2 \left(\frac{E_{er}}{\rm keV}\right)^{a_3}\right]
    \frac{\ln{\left[1 + a_0 \left(\frac{E_{er}}{\rm keV}\right)\right]}}{E_{er}},
    \label{eq:mod-thomas-imel}
\end{equation}
where $E_{er}$ is the energy of the ER, and $a_0 = 2.41 \pm 0.58$, $a_1 =  21.0\pm2.2$, $a_2 = 0.11\pm0.03$, and $a_3 =1.71\pm0.08$ are the parameters of the model constrained during the calibration measurements \cite{DarkSide:2021bnz}.\footnote{We use the following parameterization: $a_0 = \gamma\rho\:{\rm keV}$, $a_1 = 1/\gamma$, $a_2 = p_0$, and $a_3 = p_1$, where $\gamma$, $\rho$, $p_0$  and $p_1$ are the parameters of the extended Thomas-Imel box model defined in Ref.~\cite{DarkSide:2021bnz}.}
The NR ionization yield is given by
\begin{equation}
    Q_y^{NR} = 
    \frac{\left[1-f(C_{box}^{NR}, E_{nr})\right] N_i(f_B, E_{nr})}{E_{nr}},
    \label{eq:recomb_NR}
\end{equation}
where $E_{nr}$ is the NR energy, and $f(C_{box}^{NR}, E_{nr})$ and $N_i(f_B, E_{nr})$ are two analytical functions of the energy and of $C_{box}^{NR} = (8.05 \pm 0.15)\:{\rm V/cm}$ and $f_B = (0.67 \pm 0.02)$\cite{DarkSide:2021bnz}, which are in turn the two parameters of the model constrained during the calibration measurements.

The ionization response to ER (NR) is measured down to 180 eV$_{er}$ (500 eV$_{nr}$), corresponding to $\sim 9$ (3) ionization electrons, respectively. The ionization response to both ER and NR is the lowest threshold ever reached in liquid argon. These results are obtained by fitting the specific ER and NR ionization models to $\ce{^{241}Am^9Be}$ and $\ce{^{241}Am^{13}C}$ neutron sources data, $\beta$-decay data of $^{39}$Ar, and electron captures of $^{37}$Ar obtained during the \DSfs\ calibration campaign, and by external datasets from the SCENE \cite{SCENE:2014iyj} and ARIS \cite{Agnes:2018mvl}%\cite{Joshi:2014fna}
experiments.

\section{The Bayesian Network method for DarkSide-50}
\label{sec:detector-effects_responsemodel}

In this section we discuss the implementation of the BN method for the \DSfs\ response model. In particular, we describe how to keep the dependence on the calibration parameters up to the final $N_{e^-}$ spectrum, where $N_{e^-}$ is the number of reconstructed primary electrons.

The detector response can be schematized in the following consecutive steps:
\begin{enumerate}
    \item conversion of the deposited energy to a certain number of detectable quanta (e.g. number of primary electrons) produced during the interaction;
    \item detection efficiency or non linearity effects that depend on the position of the events inside the TPC;
    \item resolution effects of the photodetectors (e.g. PMTs);
    \item effects induced by trigger and analysis event selection.
\end{enumerate}
All these steps contribute to distorting the original theoretical energy spectrum into a different observed one.

\subsection{From energy deposit to reconstructed electrons after recombination} 
\label{subsec:fromkevtone}
The incoming particles interacting with the active volume inside the TPC can produce an ER or a NR. In terms of detector response, the difference in the deposited energy between the two cases lies in the production mechanisms of detectable quanta.
This dependence is described by the ionization yield of an ER or a NR.
The average number of primary ionization electrons produced is given by
\begin{equation}
    \left< N_{e^-}^{k} \right>= E_{k} Q_y^{k} \left( E_{k}, \boldsymbol{\theta}_{\rm cal} \right),
\label{eq:average_det_quanta_nr}
\end{equation}
where $k$ denotes the ER or NR process, $E_{k}$ is the energy deposited in the process $k$, $Q_y^{k}$ is the $k$ ionization yield per unit of energy and $\boldsymbol{\theta_{cal}}=\{a_0,\,a_1,\,a_2,\,a_3,\,C_{box}^{NR},\,f_B\}$ is the complete list of calibration parameters that regulates the ER or NR yield functions.
The production of an ion-electron pair surviving recombination is a stochastic process, and it can be represented by a binomial process. The maximum number of electrons that can be produced at a given energy $E_{NR}$ can be estimated as
\begin{equation}
    N_{e^-}^{{\rm max},\:{NR}} = \frac{E_{nr}}{w},
\end{equation}
where $w = (19.5\pm1.0)\:{\rm eV}$ is the liquid argon average work function~\cite{Doke:2002oab}.
It follows that the probability that a certain amount of energy is released in the TPC in the form of an ion-electron pair can be estimated, using Eq. \eqref{eq:average_det_quanta_nr}, as

\begin{equation}
    \epsilon^{NR} =\frac{\langle N_{e^-}^{NR}\rangle}{N_{e^-}^{{\rm max},\:{NR}}} = w \: Q_y^{NR} \left( E_{nr}, \boldsymbol{\theta}_{\rm cal} \right).
\end{equation}
As a result, the probability of having produced a certain number $N_{e^-}$ of ion-electron pairs is given by the binomial distribution
\begin{equation}
    P(N_{e^-}^{NR}| E_{nr}, \boldsymbol{\theta}_{\rm cal}) = \mathcal{B}(N_{e^-} \,|\, p = \epsilon^{NR}, n = N_{e^-}^{max}),
\label{eq:NR-to-ne-probability}
\end{equation}
where
\begin{equation}
    \mathcal{B}(k\:|\:p, N) = \binom{N}{k} p^k (1-p)^{N-k}.
\end{equation}

The probability of having a certain number $N_{e^-}$ of primary ionization electrons can be represented as
\begin{equation}
    P(N_{e^-}^{ER}| E_{er}, \boldsymbol{\theta}_{\rm cal}) = \int_{N_{e^-}^{ER}}^{N_{e^-}^{ER}+1} \mathcal{N}\left(x| \mu, \sigma \right)dx
\label{eq:ER-to-ne-probability}
\end{equation}
where $\mathcal{N}$ is a normal distribution with $\mu=\left< N_{e^-}^{ER} \right>$, $\sigma~=~\sqrt{F \left< N_{e^-}^{ER} \right>}$ and $F$ is the Fano factor~\cite{Fano:1947zz}. In this way the statistical fluctuations of the number of produced detectable quanta for ERs are implemented as normal fluctuations.

The expected spectrum in $N_{e^-}$ after recombination can be computed from the probabilities in Eqs.~\eqref{eq:NR-to-ne-probability} and~\eqref{eq:ER-to-ne-probability}.
Defining $x_j^{th}$ as the energy of the $j$-th bin and $y_j^{th}$ the content of the $j$-th bin of the histogram of the theoretical ER or NR spectrum, we introduce the probability matrix $\mathcal{M}^1$ as
\begin{equation}
\begin{split}
    \mathcal{M}_{i,j}^{1,\:ER} &= P(N_{e^-}^{ER} = N_{e^-}^i| E_{er} = x_j^{th}, \boldsymbol{\theta}_{\rm cal}) \\
    \mathcal{M}_{i,j}^{1,\:NR} &= P(N_{e^-}^{NR} = N_{e^-}^i| E_{nr} = x_j^{th}, \boldsymbol{\theta}_{\rm cal})
\end{split}
\label{eq:M1_definition}
\end{equation}
where $N_{e^-}^i$ is the number of primary ionization electrons in the expected spectrum. As a consequence, the expected spectrum $S_i^{N_{e^-}}$ can be computed as the product of the $\mathcal{M}^1$ matrix and the theoretical spectrum
\begin{equation}
    S_{i}^{N_{e^-}} = \sum_j \mathcal{M}^1_{i,j} y_j^{th},
\label{eq:yexpected}
\end{equation}
where we have omitted the $ER$ and $NR$ indices to simplify the notation.

The $\mathcal{M}^1$ probability matrix has some important properties. It is not a square matrix, and, since the total probability is one, the sum of the columns of the matrix is equal to unity
\begin{equation}
    \sum_{i=0}^\infty \mathcal{M}^1_{i,j} = 1 \qquad \forall \:j.
\end{equation}

What is remarkable is that the $\mathcal{M}^1$ matrix as defined in Eq.~\eqref{eq:M1_definition} explicitly depends on the calibration parameters, and thus their uncertainties play the role of the systematic uncertainties of our experiment. Taking into account these parameters directly inside the likelihood, would allow treating the systematic uncertainties in a very clean and straightforward way, without using multi-templates methods to propagate them. 
This will be clearer and more explicit in the next sections, where the analysis will be described in detail.

A different way to proceed must be considered for the Migdal effect because electrons both in the NR and ER channels are produced. Therefore, the total number of electrons produced in a single event is the sum of the electrons released by the NR and those released by the ER. The probability of releasing $N_{e^-} = N_{e^-}^{NR}+N_{e^-}^{ER}$ primary ionization electrons, given a NR with energy $E_{nr}$ and a Migdal emission depositing an energy $E_{er}$, is given by 
\begin{align}
    \small
    &p(N_{e^-}) = \nonumber\\
    &\;\sum_{N_{e^-}^{NR}} \sum_{N_{e^-}^{ER}} \sum_{E_{nr}} \sum_{E_{er}} p(N_{e^-}^{NR}, N_{e^-}^{ER} | E_{nr}, E_{er}) \times \nonumber \\
    &p(E_{nr}, E_{er})\delta_{N_{e^-}, N_{e^-}^{NR}+N_{e^-}^{ER}} =\nonumber \\
    & \sum_{N_{e^-}^{NR} = 0}^{N_{e^-}} \sum_{E_{nr}} \sum_{E_{er}} p(N_{e^-}^{NR}| E_{nr}) p(N_{e^-} -  N_{e^-}^{NR}| E_{er})\times \nonumber \\
    &p(E_{nr}, E_{er}),
\end{align}
where we did not report the dependence on $\boldsymbol{\theta}_{cal}$ for shortness of notation. In the above equation, which is valid assuming no interference between the two energy releases during recombination, we can identify $p(N_{e^-}^{NR} = N_{e^-}^k| E_{nr} = x_j^{th})$ with $\mathcal{M}^{1,NR}_{k,j}$, see Eq.~\eqref{eq:M1_definition}. Furthermore, $p(N_{e^-}^{ER} = N_{e^-}^i - N_{e^-}^k| E_{er} = x_l^{th})$ is equal to $\mathcal{M}^{1,ER}_{i-k,l}$, see Eq.~\eqref{eq:M1_definition}, and $p(E_{nr} = x_j^{th}, E_{er} = x_l^{th})$ is the double differential Migdal rate normalized to unity.
Making the indices explicit, we obtain 
\begin{equation}
\begin{split}
    &p(N_{e^-} = N_{e^-}^i) = \\
    & \sum_{k = 0}^{i} \left( \mathcal{M}^{1,NR} \cdot \frac{d^2\tilde{R}_M}{dE_{er}dE_{nr}} \cdot
    \left(\mathcal{M}^{1,ER}\right)^T \right)_{k, i-k},
    \\
    \label{eq:migdal_prob}
\end{split}
\end{equation}
where $\tilde{R}$ indicates the normalized double differential Migdal rate,  satisfying the condition
\begin{equation}
    \sum_{j} \sum_{l} \frac{d^2\tilde{R}_M}{dE_{er}dE_{nr}}\Bigg|_{j, l} = 1.
\end{equation}
The Migdal expected spectrum $S_i^{N_{e^-}, Mig}$ after recombination is simply obtained by substituting the normalized double differential Migdal rate in Eq.~\ref{eq:migdal_prob} with the original one.

\subsection{Detector effects: efficiency, radial corrections, and PMT response}
\label{sec:detector-effects}
The two main effects that play a sizable role in the reduction of the S2 signal yield are the so-called radial correction and electron lifetime efficiency~\cite{DarkSide:2021bnz}. The former effect can be parameterized as an efficiency factor depending only on the radial position of the event. The latter is instead due to the electron lifetime inside the TPC, namely the probability that an ionization electron is captured by electro-negative impurities in the liquid chamber. This is a small effect that depends mainly on the distance between the event and the liquid-gas interface, and therefore on the event depth inside the TPC.

Since we are dealing with energy spectra where the information on the position of the events has been integrated out, we simulate the theoretical spectra retaining the energy and 3D position information.
As a consequence, we have access to the position and the related total correction factor of each event. We use this information to numerically compute the \emph{pdf} to have a certain correction $\epsilon_{src}^{ch}$ 
in a given event. 
This \emph{pdf} depends on the PMT channel and on the background/signal source spatial features. The result of this procedure is a set of numerical functions (one per source and PMT channel) over which we integrate to get the full convoluted effect. These \emph{pdf}s depend only on the theoretical spectra, and do not depend on the calibration parameters. They can be therefore computed once before the final MCMC integration, and used as an input for the analysis.

An additional effect due to the PMT channel corrections has to be taken into account. In fact, the contributions from the various PMT channels are not simply summed up in the data spectrum. The PMT response has been equalized to the central PMT. As a consequence, we need to apply the same correction when implementing the detector response model to be compared with data. This correction is a simple multiplicative factor $r_{ch}$ in the PMT energy response that depends on the PMT channel.

Finally, the PMT resolution effect can be treated as a Gaussian smearing effect, similar to the ER Fano fluctuations of Eq.~\eqref{eq:ER-to-ne-probability}. We can therefore express the probability that an event from the source $src$ in the PMT channel $ch$ is detected to have $N_{e^-}^f = i$ extracted electrons having originally produced $N_{e^-}^{(0)} = j$ electrons surviving recombination as
\begin{equation}
\begin{aligned}
    &P(N_{e^-}^f = i | N_{e^-}^{(0)} = j, ch, src) = \\
    &\;\int_{i}^{i+b_w} \;dx \int d\epsilon_{src}^{ch} \,\mathcal{N} \left(x | \mu, \sigma \right) p(\epsilon_{src}^{ch}),
\label{eq:Ne-to-Ne-det_response}
\end{aligned}
\end{equation}
where $\mu = \epsilon_{src}^{ch} r_{ch} j$, $\sigma = S \sqrt{\epsilon_{src}^{ ch} r_{ch} j}$, $b_w$ is the width of the bins of the final observed $N_{e^-}$ histogram, $\epsilon_{src}^{ch}$ is the overall correction factor and $p(\epsilon_{src}^{ch})$ is its \emph{pdf} as computed by means of the Monte Carlo. Furthermore, $\mathcal{N}(x|\mu, \sigma)$ is the normal \emph{pdf} with mean $\mu$ and variance $\sigma$, $r_{ch}$ is the channel correction factor, and $S = 0.27$ is a width factor that has been determined during the calibration. We then define a set of matrices
\begin{equation}
    \mathcal{M}^2_{i, j}(ch, src) = P(N_{e^-}^f = i | N_{e^-}^{(0)} = j, ch, src)
\label{eq:M2_definition}
\end{equation}
such that we can compute the final expected $N_{e^-}$ spectrum $S^f_{ch, src}$ in the PMT channel $ch$ induced by the source $src$ as
\begin{equation}
   S^f_{ch, src, i} (\boldsymbol{\theta}_{cal})= \sum_{j = 1}^{N_{max}^{exp}} \mathcal{M}^2_{i, j} (ch, src) S^{N_{e^-}}_{ch, j} (\boldsymbol{\theta}_{cal}).
\end{equation}
The $\mathcal{M}^2$ matrix depends only on the channel corrections, on the width factor $S$ and, by means of the \emph{pdf}, on the theoretical spectra. These quantities do not depend on the calibration parameters $\boldsymbol{\theta}_{cal}$. They can be computed once before the final MCMC integration, and used as an input for the analysis. This gives great benefits in terms of performance, because it means that, unlike the two $\mathcal{M}^1$ matrices, $\mathcal{M}^2$ does not need to be computed at each step of the MCMC algorithm.

In conclusion, once we have $S^f_{ch, src}$, we can obtain the final expected spectrum $S^f_{ER, NR}$ by summing over all the PMT channels and over all the sources
\begin{equation}
\begin{split}
    S^f_{NR, i} (\boldsymbol{\theta}_{cal}) &= \sum_{N_{ch}} S^f_{ch, NR} (\boldsymbol{\theta}_{cal}), \\
    S^f_{ER, i} (\boldsymbol{\theta}_{cal}) &= \sum_{N_{ch}}\sum_{N_{src}} S^f_{ch, src, ER} (\boldsymbol{\theta}_{cal}). \\
\end{split}   
\label{eq:spectra-as-function-of-det-par}
\end{equation}

We validated the new NR and ER response model on the implementation of code used for the calibration, for the published DS-50 and for the Migdal analyses~\cite{DarkSide:2021bnz,DarkSide-50:2022qzh,DarkSide:2022dhx}, and based on a toy MC simulation of the detector response.

\section{Likelihood and parameters}
\label{sec:likelihood}

\begin{figure}
    \centering
    \includegraphics[width = 0.5\textwidth]{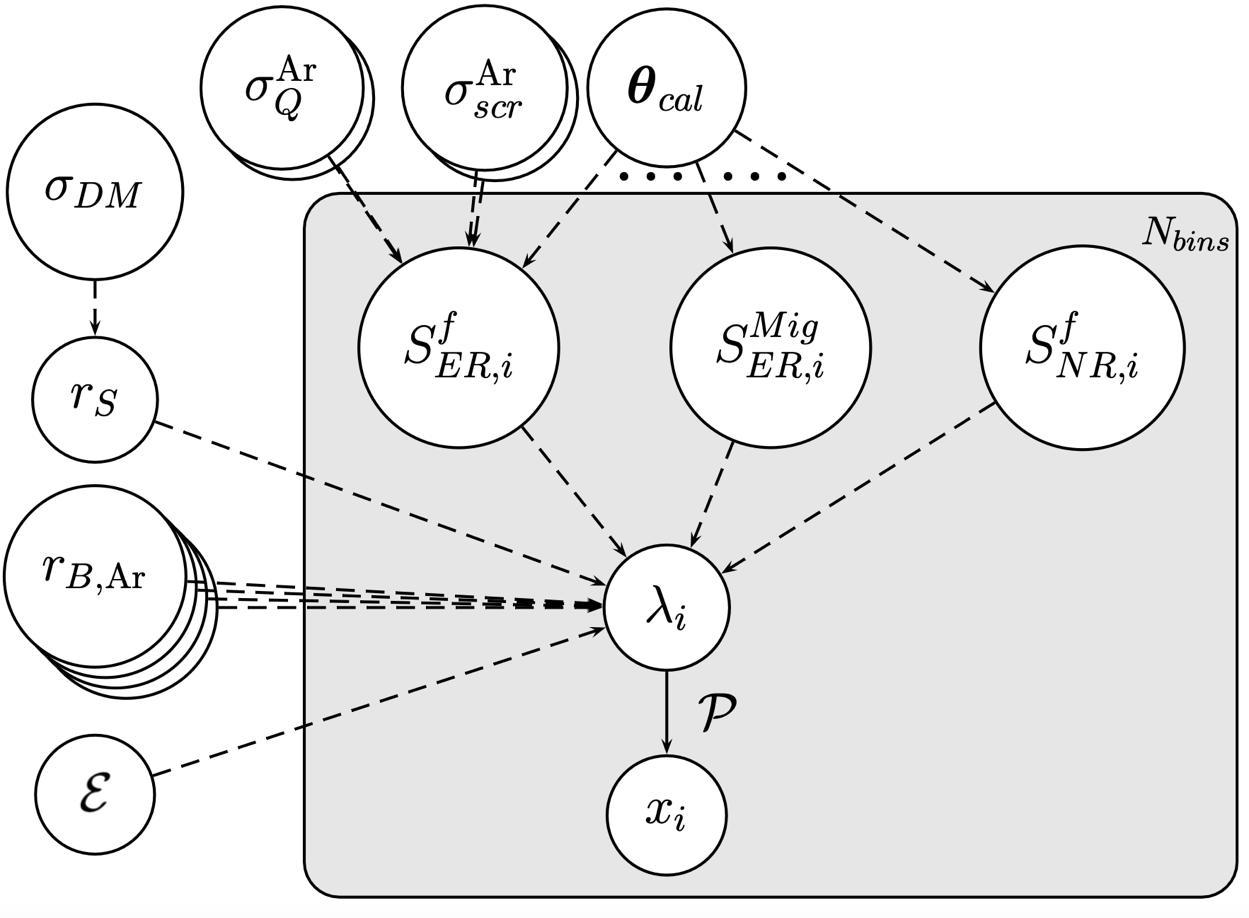}
    \caption{\label{fig:Complete-Network} Bayesian network for the \DSfs\ low-mass analysis: the node $x_i$ is the observed number of events in the $i$-th bin; $\lambda_i$ is the intensity of the Poisson process in the $i$-th bin; the gray box contains the nodes and connections that are repeated over the total number of bins $N_{bins}$; the dots represent the portion of the network implementing the detector response model in the form of the two smearing matrices $\mathcal{M}^1$ and $\mathcal{M}^2$; besides $\sigma_{\rm DM}$, which is directly related to $r_S$ in our case, all the nodes outside the gray box are the free parameters of the likelihood, as described in Sec.~\ref{sec:likelihood}.}
\end{figure}

In the following, we assume a bin-by-bin Poisson likelihood defined as
\begin{equation}
    p(\{x_i\} | \boldsymbol{\theta}) = \prod_i
    \frac{\lambda_i(\boldsymbol{\theta})^{x_i}}{x_i !} e^{-\lambda_i(\boldsymbol{\theta})},
    \label{eq:likelihood}
\end{equation}
where $x_i$ denotes the number of events in the data spectrum in the $i$-th bin, $\boldsymbol{\theta}$ indicates generically all the parameters of the fit - i.e. the calibrations, the signal rate, and the background rates. In the case under consideration, $\lambda_i(\boldsymbol{\theta})$ is given by
%\begin{equation}
\begin{align}
    \lambda_i =\frac{\mathcal{E}}{\mathcal{E_{\rm DS}}} \:\Big[&r_{B, {\rm Ar}} \; S_i^{\rm Ar}(\boldsymbol{\theta}_{cal}, \sigma_{scr}^{\rm Ar}, \sigma_Q^{\rm Ar}) + \nonumber \\
    &r_{B, {\rm Kr}} \; 
    S_i^{\rm Kr}(\boldsymbol{\theta}_{cal}, \sigma_{scr}^{\rm Kr}, \sigma_Q^{\rm Kr}) +\nonumber\\
    &r_{B, {\rm PMT}} \; S_i^{\rm PMT}(\boldsymbol{\theta}_{cal}) +\nonumber \\ 
    &r_{B, {\rm cryo}} \; S_i^{\rm cryo}(\boldsymbol{\theta}_{cal})  +\nonumber \\
    &r_{S} \; \Big( S_i^{\rm NR}(\boldsymbol{\theta}_{cal}) 
    + S_i^{\rm Mig}(\boldsymbol{\theta}_{cal}) \Big) \Big]. 
    \label{eq:lambda}
\end{align}
%\end{equation}
This parametrization of $\lambda$ is an extension to the one exploited in Ref.~\cite{GrillidiCortona:2020owp} to include the detector response model.
Here $S_i^{src}$ represent the expected background and signal spectra for the total exposure $\mathcal{E_{\rm DS}}$ as a result of the detector response, see Eq.~\eqref{eq:spectra-as-function-of-det-par}.
The variables $r_{B, src}$ are proportional to the rate of the internal and external background components. Since, for simplicity, the background spectra are normalized to the \DSfs\ exposure, they are normalized such that $r_{B, src} = 1$ corresponds to the case in which the exposure is equal to the \DSfs\ exposure.

The parameter $r_{S}$ is proportional to the signal rate with a flat prior. The signal spectra are normalized to the \DSfs\ exposure and computed for $\sigma_{SI}^{DM} = 10^{-38}\:{\rm cm}^2$. Therefore, $r_S = 1$ corresponds to $\sigma_{SI}^{DM} = 10^{-38}\:{\rm cm}^2$.
The total exposure is denoted by $\mathcal{E}$ and $\mathcal{E}_{\rm DS}$ is the \DSfs\ exposure of the analyzed dataset. The uncertainty on the exposure is implemented with a normal prior with a $1.5\%$ standard deviation~\cite{DarkSide-50:2022qzh}.
The various $S_i^{src}$ spectra are in turn functions of a possibly different subsets of parameters encoding various aspects of the detector response and background model, and spectrum-specific uncertainties. 

The complete likelihood is represented as a Bayesian Network in Fig.~\ref{fig:Complete-Network}. As we discussed in Sec.~\ref{sec:concept}, this representation makes clear the structure of Eq.~\eqref{eq:likelihood}--\eqref{eq:lambda} and the dependence of the spectra on the various nuisance parameters.  
In the following, we give a detailed description of its implementation.

\subsection{Signal spectra}
The signal spectra are explicitly described with two contributions: one due to NR interactions $(S^{\mathrm{NR}})$, and another one describing the Migdal effect $(S^{\mathrm{Mig}})$. The nuclear recoil contribution is derived assuming the Standard Halo Model with a DM escape velocity $v_{esc}=544$ km/s, the local standard at rest velocity $v_0=238$ km/s, $v_{Earth}=232$ km/s, and the DM density $\rho_{DM}=0.3$ GeV/c$^2$/cm$^3$~\cite{Baxter:2021pqo}. The Migdal effect contribution to the signal spectra has been computed as in Ref.~\cite{DarkSide:2022dhx} exploiting the Migdal probabilities computed in Ref.~\cite{Ibe:2017yqa}. 

\begin{figure}
    \centering
    \includegraphics[width = 0.49\textwidth]{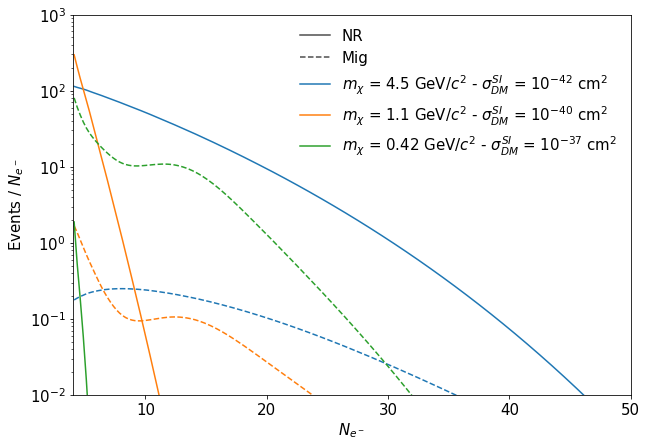}
    \caption{Signal events in the $N_{e^-} = \left[4, 50\right]$ range assuming the \DSfs\ exposure. The colors represent different DM masses for the NR (solid) and Migdal (dashed) processes.}
    \label{fig:signal-spectrum}
\end{figure}

Figure~\ref{fig:signal-spectrum} shows the signal spectra for DM masses of $4.5\,{\rm GeV}/c^2$ (blue), $1.1\,{\rm GeV}/c^2$ (orange) and $0.42\,{\rm GeV}/c^2$ (green), for $\sigma_{\mathrm{DM}}^{\mathrm{SI}} = 10^{-38}\,\mathrm{cm}^2$. The solid curves show only the NR contribution, while the dashed ones only the Migdal effect. 

We are not considering any theoretical systematic uncertainty on the signal spectra, nevertheless they could be easily implemented as additional nuisance parameters. 

\subsection{Background spectra}
\begin{figure}
    \centering
    \includegraphics[width = 0.49\textwidth]{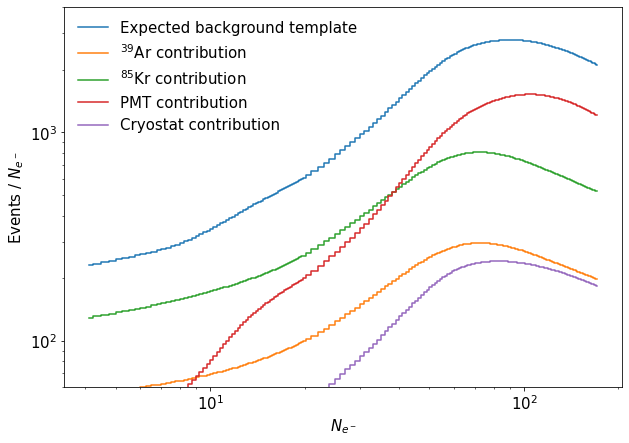}
    \caption{The so-called Asimov dataset, namely the expected background (blue line) in the $N_{e^-} = \left[4, 170\right]$ range, assuming the \DSfs\ exposure. The different colored lines represent the expected contributions from the different background sources. The spectra have a $0.25$ $N_{e^-}$ binning for $N_{e^-} < 20$ and a $1$ $N_{e^-}$ elsewhere, see text. }
    \label{fig:pseudodata}
\end{figure}

The background contributions described in Sec.~\ref{sec:bkgd_model} and shown in Fig.~\ref{fig:pseudodata} are separated in four main spectra, two internal induced by $^{39}$Ar and $^{85}$Kr, and two external produced by the materials of the PMT and the cryostat. They are denoted respectively with $S^{\rm Ar}$, $S^{\rm Kr}$, $S^{\rm PMT}$, and $S^{\rm cryo}$.  
The relative normalizations have been determined by dedicated measurements and Monte Carlo simulations, as described in Ref.~\cite{DarkSide-50:2022qzh}. Their systematic uncertainties are implemented as a normal prior centered in 1 and with a $14\%$ ($\ce{^{39}{\rm Ar}}$), $4.7\%$ ($^{85}{\rm Kr}$), 
    $12.6\%$ (PMT) and $6.6\%$ (cryostat) standard deviation.

The ${\rm^{39}\!Ar}$ and the ${\rm^{85}\!Kr}$ background spectra are affected by screening effects and Q-value uncertainties parametrized with $\sigma_{scr}^{\rm Ar, Kr}$ and $\sigma_{Q}^{\rm Ar, Kr}$, respectively. 
The Q-value uncertainties are implemented as two Gaussian parameters regulating the $\ce{^{39}{\rm Ar}}$ and $\ce{^{85}{\rm Kr}}$ spectra, with the intensity dependent on the energy.
Specifically, the energy spectrum, for $k= \ce{^{39}{\rm Ar}},\,\ce{^{85}{\rm Kr}}$, is given by:
\begin{align}
&y_{th}^{k} (E_{er}) = y_{th}^{k, (0)}(E_{er})
\Bigg[ 1
+ r(E_{er}) \:\sigma_{Q}^{k}\\ 
&+ \left(0.1-\frac{E_{er}}{2\:{\rm keV}} \right) \sigma_{scr}^{k} \,\Theta(200-E_{er}/{\rm eV}) \Bigg] \nonumber
\end{align}
where $y_{th}^{k, (0)}(E_{er})$ is the \emph{central} theoretical spectrum, $r(E_{er})$ is the relative uncertainty due to the uncertainty on the Q-value, $\Theta$ is the Heaviside function, and $\sigma_{scr}^k$ and $\sigma_{Q}^k$ are both controlled by a standardized normal \emph{pdf} $\mathcal{N}(0,1)$ as prior function.

\subsection{Spectra dependence on the detector response parameters}
\label{sec:discorsobande}

Both the signal and the background spectra depend on $\boldsymbol{\theta}_{cal}$ which represents the calibration parameters. 
The result of the calibration measurements \cite{DarkSide:2021bnz} are implemented as a multivariate normal prior.

Eq.~\eqref{eq:spectra-as-function-of-det-par} allows us to compute for any specific detector response configuration (identified by a choice of the parameters $\boldsymbol{\theta}_{cal}$) the associated background and signal spectra.
In this section, we give some examples on how the spectra change as a function of $\boldsymbol{\theta}_{cal}$ and show how our implementation is substantially different from constructing average spectra with the corresponding $\pm \sigma$ variations.

\begin{figure*}
    \centering
    \includegraphics[width = 0.288\textwidth]{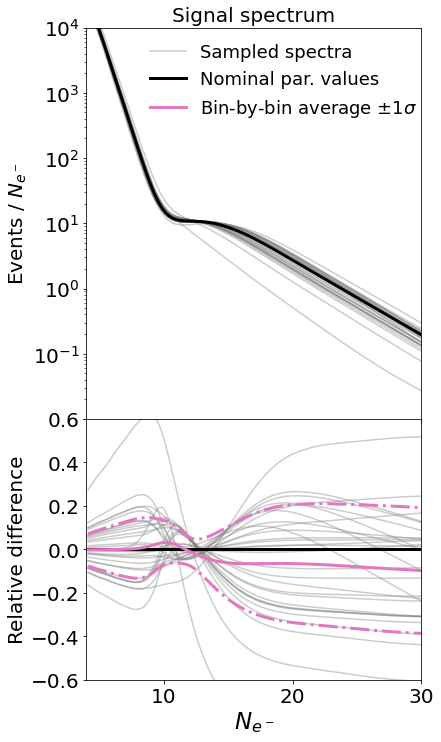}
    \includegraphics[width = 0.278\textwidth]{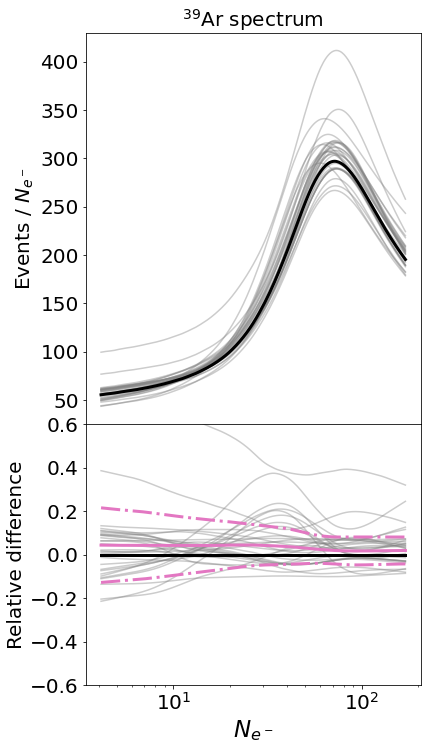}
    \includegraphics[width = 0.408\textwidth]{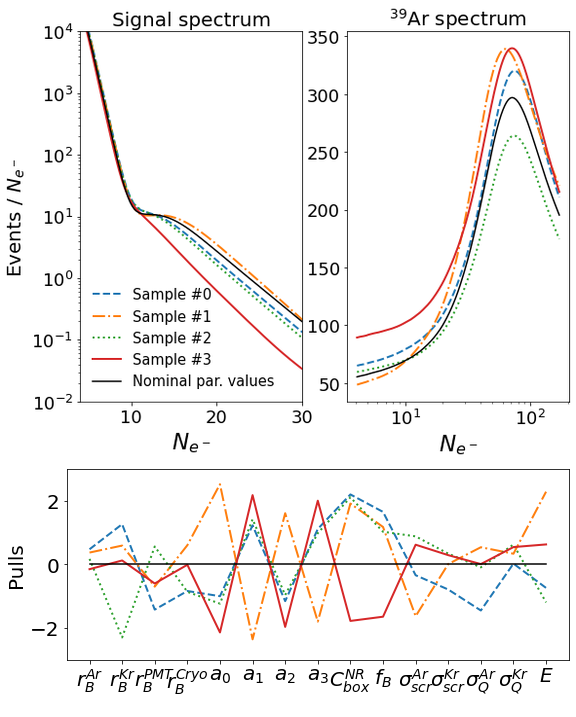}
    \caption{({\bf Left}) Signal (left) $\ce{^{39}Ar}$ background (right) spectra obtained by randomly selecting 30 points from the nuisance parameters \emph{pdf}, the bottom plots show the relative difference with the spectrum obtained with the nominal parameter values (black solid line) and the average (pink solid curve) $\pm\sigma$ envelopes (pink dashed lines) calculated bin by bin. ({\bf Right}) Four different signal (left) and $\ce{^{39}Ar}$ background spectra (right) showing significant shape differences with respect to the nominal spectrum, which is also shown; the bottom plot shows the nuisance parameters pulls for each of the spectra selected above.  
    }
    \label{fig:bkg-spectra-syst}
\end{figure*}
Figure~\ref{fig:bkg-spectra-syst} shows the effect of sampling from the prior \emph{pdf} of the nuisance parameters onto the signal and  the $\ce{^{39}Ar}$ spectra. We chose the $\ce{^{39}Ar}$ as a representative background. We show only 30 different randomly selected spectra to keep the plot readable. We stress that these spectra are not chosen on the basis of their shapes. The bottom part of each sub-figure shows the relative difference with respect to the spectrum obtained keeping all nuisance parameters to the central value of their prior (nominal spectrum).
From these figures, it is visible how the nuisance parameters have a non-trivial effect on the spectra: (1) overall as well as in specific bins there is a sizable difference with respect to the reference spectrum; (2) more importantly, there is a significant shape difference, possibly with more than one crossing.

At any given value of $N_e$ we can construct the \emph{pdf} of the expected number of events and compute the expected value and the $\pm \sigma$ intervals. From these values we can construct the average spectrum (solid pink curve), and the $\pm \sigma$ (dash-dotted pink curve).
Given the non-linearity of the problem, in general the envelope spectra are different from those computed with the corresponding choice of the $\boldsymbol{\theta}_{cal}$ parameters.
Sizable differences are clearly visible both in total event yield and in shape, especially for the signal spectrum.
In addition, the envelope spectrum loses part of the bin-bin correlations, and it may represent a spectrum which does not correspond to any specific configuration $\tilde{\boldsymbol{\theta}}_{cal}$ of the detector response.
The method presented here naturally takes into account the correlations between the $\boldsymbol{\theta}_{cal}$ parameters,\footnote{No assumption on $\boldsymbol{\theta}_{cal}$ is needed to produce the $\pm\sigma$ spectra because their joint \emph{pdf} are included in the method.} and the correlation between different spectra induced by the $\boldsymbol{\theta}_{cal}$ parameters.

The right plots of Fig.~\ref{fig:bkg-spectra-syst} show four different signal and $\ce{^{39}Ar}$ spectra that manifest large differences with respect to the nominal spectrum (black curve). For each of these spectra, the bottom plot shows the nuisance parameters pulls with respect to their prior expected values. Although these configurations are not very likely, $2\,\sigma$ variations in the parameters' value have a substantial impact on the spectrum, inducing effects as large as doubling the event yield in certain bins.

The implementation in the signal and background models of the systematic variation of the detector response described in this section is the key element for the correct uncertainty propagation in the fit results.

\section{Fitting procedure and code implementation}
\label{sec:fitproc}
The fit samples from the space of possible detector response models, computes the corresponding spectra both for the signal and backgrounds, tries to fit the observed spectrum, and finally weights the result by the probability of the specific response model sampled. In this way the average model and the corresponding uncertainty are computed from the \emph{pdf} of the model that fits the observed spectra.

The posterior of all the parameters is sampled by means of the Metropolis-Hastings MCMC algorithm, as implemented in {\tt BAT}~\cite{2009CoPhC.180.2197C,ref:BAT}. The {\tt BAT} package is a set of {\tt C++} libraries implementing statistical tools for Bayesian analyses, that has been largely used in experimental and phenomenological analyses.

Since the mathematics behind each single step of the chain - i.e. how the calibration parameters are connected to the final spectra - is mostly linear algebra, this part of the code has been implemented on GPUs by means of the {\tt CUDA} libraries~\cite{cuda}.
A generic implementation of this method is provided in a public {\tt GitHub} repository \cite{github}.

In all the following analyses, we used the standard {\tt BAT} tools to guarantee that the Markov chains are stable and accurate.
In particular, the convergence is obtained by means of the {\tt BAT} pre-run phase that assures by tuning the Metropolis-Hastings MCMC parameters that all the parallel chains converge to the same region of the parameters' phase space with an optimal Metropolis-Hastings MC rejection rate. In order to achieve the required statistical accuracy, the sampling is performed by means of 12 parallel MCMC chains, for a total number of steps equal to $1.2 \times 10^6$.

From the implementation point of view, the background model or background-plus-signal model are implemented as an extension of the {\tt BAT} {\tt BCModel} class. The likelihood and the priors are customized following the prescriptions of the previous sections, and the linear algebra implemented in CUDA. The inputs of the fit, i.e. the theoretical spectra and the $\mathcal{M}^2$ smearing matrices, are read at the constructor level. The GPU preliminary operations are also performed once inside the constructor, to optimize the sampling procedure. A complete fit, including also the prerun phase, takes $4.8\:{\rm h}$ ($40\:{\rm min}$ for prerun) with 12 treads on an Intel Xeon Silver 4216 CPU ($2.1\:{\rm GHz}$) and a Nvidia GeForce RTX 3090 GPU. This roughly corresponds to a sampling rate of $6.8\:{\rm Hz}$ per chain.

\section{Results}
\label{sec:results}

We compute the upper bound for the DM signal as the 90\% Credible Interval (C.I.). This is defined as the value of $\sigma_{SI}^{DM}(m_{DM})$ corresponding to the 90\% quantile of the marginalized posterior \emph{pdf} for $r_S$.

In order to avoid possible biases coming from using data, we performed a blind analysis based on a pseudo-dataset. This dataset has been generated from the expected background template, the so-called Asimov dataset, which is plotted in Fig.~\ref{fig:pseudodata}. 

If not specified otherwise, all the fits are performed in the full $N_{e^-} = \left[4, 170\right]$ range. Below $N_{e^-} = 4$, the data are indeed dominated by correlated events, mostly due to the contamination of spurious electrons~\cite{DarkSide-50:2022qzh}. On the other hand, the upper threshold is chosen as the maximum point at which the detector response calibrations have been validated, namely $N_{e^-} = 170$~\cite{DarkSide:2021bnz}.

The binning of the observed spectrum is denser ($0.25$ $N_{e^-}$ binning) in the $N_{e^-}<20$ region with respect to the higher $N_{e^-}$ region ($1$ $N_{e^-}$ binning).
Since the DM signal is exponentially falling, having a denser binning in the low $N_{e^-}$ region has a beneficial 5 – 10\% impact on the sensitivity in the whole mass region explored by our analysis.

\subsection{Impact of the systematic uncertainties on the limit}

As a first study, we investigate the impact of the systematic uncertainties on the final limit. For simplicity, we focus on a single DM mass value $m_{DM} = 4.5\,{\rm GeV}/c^2$, in which the Migdal effect contribution to the signal spectrum is negligible. We arranged the nuisance parameters in the following 3 groups
\begin{enumerate}
    \item group CAL: $a_0$, $a_1$, $a_2$, $a_3$, $C_{box}^{NR}$, $f_B$, which represent the calibration parameters;
    \item group TH: $\sigma_{scr}^{\rm Ar, Kr}$, $\sigma_Q^{\rm Ar, Kr}$, which represent systematic uncertainties on the theoretical background spectra;
    \item group NORM: $E$, $r_{B, src}$, which represent the systematic uncertainties on the normalization of the background spectra.
\end{enumerate}
We therefore performed 6 fits, with the following assumptions:
\begin{enumerate}
    \item all the systematic parameters are free;
    \item the group CAL is free, while all the other parameters are fixed to their expected values;
    \item the group NORM+TH are free, while all the other parameters are fixed to their expected values;
    \item the group TH is free, while all the other parameters are fixed to their expected values;
    \item the group NORM is free, while all the other parameters are fixed to their expected values;
    \item all the systematic parameters are fixed to their expected values. This corresponds to a statistic-only fit, with $r_S$ as the only active parameter.
\end{enumerate}
The results in terms of the sensitivity to the DM cross-section are given in Table~\ref{tab:limits_sys}. 
%%%%%%%%%%%%%%%%%%
\begin{table}
\small
\def\arraystretch{1.1}
\centering
\begin{tabular}{l c}
\hline
Systematics & $\sigma_{SI}^{DM} \left[90\% \:{\rm C.I.}\right] [10^{-43}{\rm cm}^2]$ \\
\hline
\hline
All              & $2.1$ \\
CAL              & $1.8$ \\
NORM + TH        & $1.7$ \\
TH               & $1.6$ \\
NORM             & $1.5$ \\
None             & $1.3$ \\
\hline
\end{tabular}
 \caption{Upper bound results on $\sigma_{SI}^{DM}$ for $m_{DM} = 4.5\:{\rm GeV}/c^2$ using the expected pseudo-dataset and fixing different groups of systematic parameters, see text.}
\label{tab:limits_sys}
\end{table}

There are two relevant differences between our innovative implementation and the published one. First of all the published approach computes the sensitivity as the frequentist 90\% C.L., while here we perform the analysis in the Bayesian approach, and we quote the $90\%$ quantile of the posterior \emph{pdf} of the signal strength parameter $r_S$~\cite{ParticleDataGroup:2020ssz}. Even if the two quantities aim at expressing the experimental sensitivity, they are conceptually different, and are defined in totally different ways. However, when operating with the same inputs, they should give numerically comparable results. In addition, when we look at the posterior of $r_S$, we are integrating over all the nuisance parameter space, while the published approach is based on the profiling of the likelihood. If the likelihood is Gaussian the marginalization and the profiling give the same result. On the other hand, in the general case in which the Gaussian assumption is not valid, the profiling procedure typically returns underestimated propagated uncertainties, and as a consequence stronger limits~\cite{Loredo:2004}.

We find that the impact of the systematic uncertainties on the calibration parameters has the same size as the impact of all the other systematic parameters, including the parameters regulating the normalization of the background and the theoretical uncertainties on the input spectra, and therefore the correct propagation of their uncertainty is relevant in the final result.
For completeness, we report in Fig.~\ref{fig:4.52_expected_posterior} in~\ref{app:exp_full_pos} the full multidimensional posterior \emph{pdf}s of the complete fit, in which all the systematic parameters are not fixed. 

As a further crosscheck we also studied the dependence of the parameters' mean and standard uncertainty of the posterior \emph{pdf}s as a function of the DM mass $m_{DM}$: none of the parameters shows a strong correlation with the DM mass, demonstrating the stability of the fitting procedure.

%================================================================================
\subsection{Impact of the high \texorpdfstring{$N_{e^-}$}{Ne} data points on the limit and the calibration parameters}

In this subsection we show the impact of choosing different $N_{e^-}$ windows on the calibration parameters' posterior \emph{pdf}s, and, as a consequence, on the upper limit on the DM cross-section.
Since the priors on the calibration parameters correspond to the constraints from the calibration, the differences between priors and posteriors tell us that the data give additional information with respect to the calibration measurements. A better knowledge on the calibration parameters allows obtaining a limit less affected by the systematic uncertainties - see the ``None'' result of Tab.~\ref{tab:limits_sys}. In order to investigate this effect, we decided to perform a fit to the expected pseudo-dataset in the regions $N_{e^-} = \left[4,30\right]$ and $N_{e^-} = \left[4,170\right]$. The long high $N_{e^-}$ tail provides better constraints on the calibration parameters.

\begin{figure*}
    \centering
    \includegraphics[width = 0.40\textwidth]{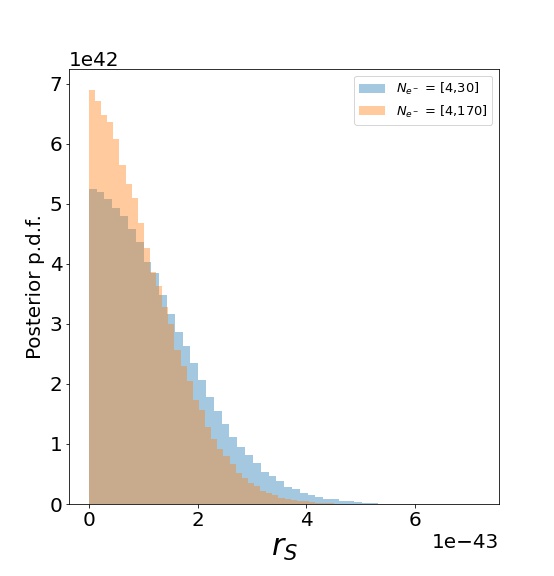}
    \includegraphics[width = 0.40\textwidth]{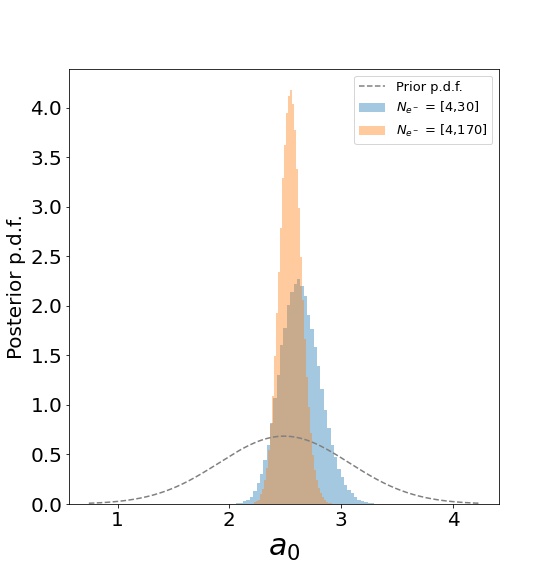}
    \includegraphics[width = 0.40\textwidth]{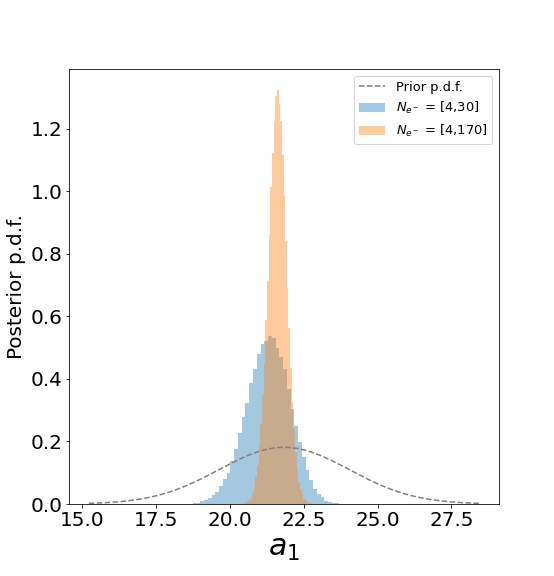}
    \includegraphics[width = 0.40\textwidth]{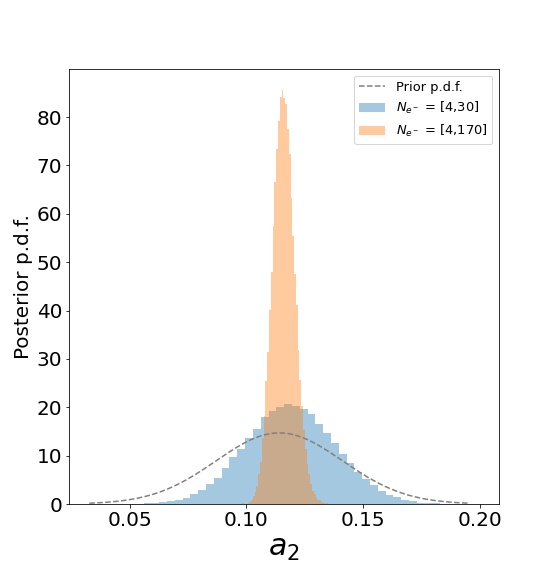}
    \includegraphics[width = 0.40\textwidth]{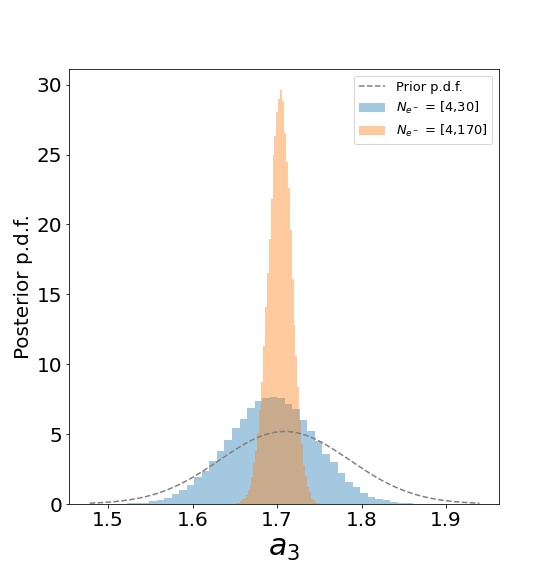}
    \includegraphics[width = 0.40\textwidth]{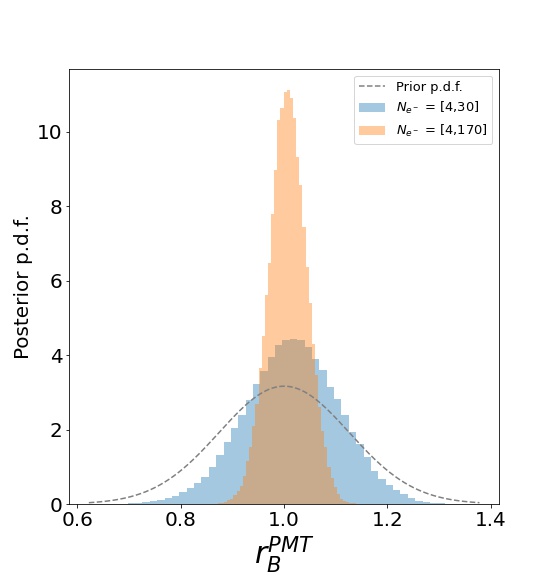}
    \caption{
    Posterior \emph{pdf}s obtained by fitting in different energy ranges a pseudo-dataset with the full background plus signal model assuming a DM mass $m_{DM} = 4.5\:{\rm GeV}/c^2$.
    The blue shaded histograms correspond to the $N_{e^-} = \left[4,30\right]$ window, the orange shaded histograms correspond to the $N_{e^-} = \left[4,170\right]$ window, and the gray dashed line are the prior \emph{pdf} of each parameter.}
    \label{fig:energy_windows}
\end{figure*}

In Fig.~\ref{fig:energy_windows} we show the results of two background plus signal fits on the pseudo-dataset performed in the regions $N_{e^-} = \left[4,30\right]$ and $N_{e^-} = \left[4,170\right]$ (orange and blue shaded histograms, respectively) with respect to the prior \emph{pdf} (gray dashed line), assuming a DM mass $m_{{DM}} = 4.5\:{\rm GeV}/c^2$.
The figure shows how the width of the calibration parameters' posteriors are smaller in the $N_{e^-} = \left[4,170\right]$ case rather than both in the prior and the $N_{e^-} = \left[4,30\right]$ cases. 
In particular, the posterior \emph{pdf}s of the ER group are more constrained than the prior \emph{pdf}s. This behavior does not depend on possible features in the data, since this study is performed on Asimov pseudo-dataset.
On the other hand, the posterior \emph{pdf}s of the NR group are similar to their priors. This is expected since in the pseudo-dataset there is no injected signal and the NR signal is relevant only in the first few bins. As a consequence, the larger data-set does not give additional information on the calibration parameters with respect to the calibration measurements. 
For this reason we obtain the best sensitivity in the $N_{e^-} = \left[4,170\right]$ case - i.e. $\sigma_{SI}^{DM} \left[90\% \:{\rm C.I.}\right] = 2.1\times 10^{-43}\:{\rm cm^2}$ - while in the $N_{e^-} = \left[4,30\right]$ the sensitivity is weaker - i.e. $\sigma_{SI}^{DM} \left[90\% \:{\rm C.I.}\right] = 2.5\times 10^{-43}\:{\rm cm^2}$.
The other parameters, not shown in the figure, do not exhibit a significant dependence on the range of the fit.

In summary, in this new implementation of the Baye\-sian approach we complement the calibration results by further constraining some of its parameters, thus reducing their uncertainties by a factor of $\sim 2$. We can directly benefit from this improvement by having a stronger bound on the signal and being able to justify it in terms of a better knowledge of the calibration parameters.

\subsection{Background-only fit on data}
\begin{figure}
    \centering
    \includegraphics[width = 0.49\textwidth]{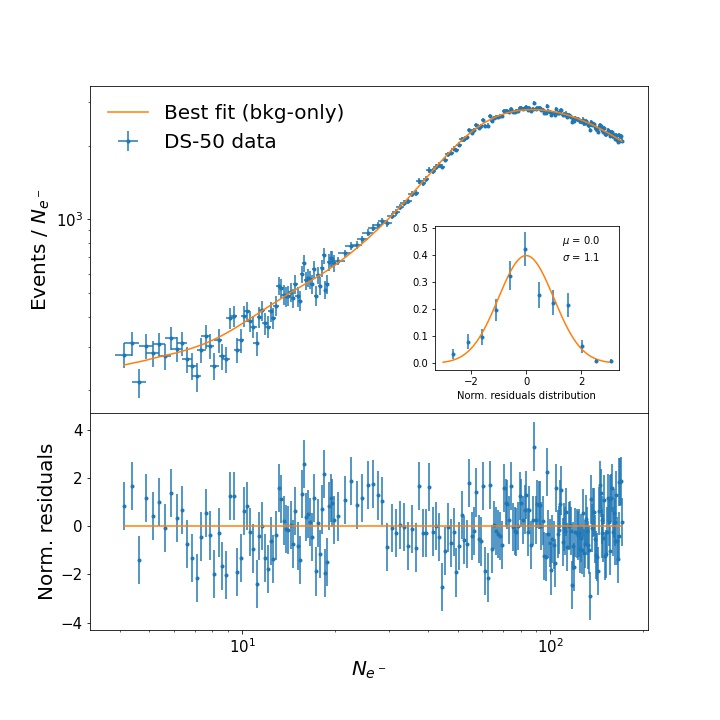}
    \caption{
    {\bf Top}: Best fit (orange line) to the observed dataset (blue points) of the \DSfs\ experiment.
    {\bf Bottom:} Normalized residual (blue points) with respect to the best fit.
    }
    \label{fig:bkgonly_fit}
\end{figure}

We perform a background-only fit on the data: the best fit (top) and the correspondent normalized residuals (bottom) are reported in Fig.~\ref{fig:bkgonly_fit}. 
The figure shows that the background model describes the observed data without noticeable deviations or excesses. The behavior of the residual is satisfactory and their distribution over all the data points is consistent with the expectation of a standard normal \emph{pdf}.

In terms of posterior \emph{pdf} on the fit parameters, no significant deviation from the prior distribution (the results of the calibration measurements) is observed. In Fig.~\ref{fig:bkgonly-posterior} in~\ref{app:bkgonly_full_pos} we report the full multidimensional posterior of the fit.

\subsection{Expected and observed limit of the DarkSide-50 experiment}

\begin{figure}
    \centering
    \includegraphics[width = 0.49\textwidth]{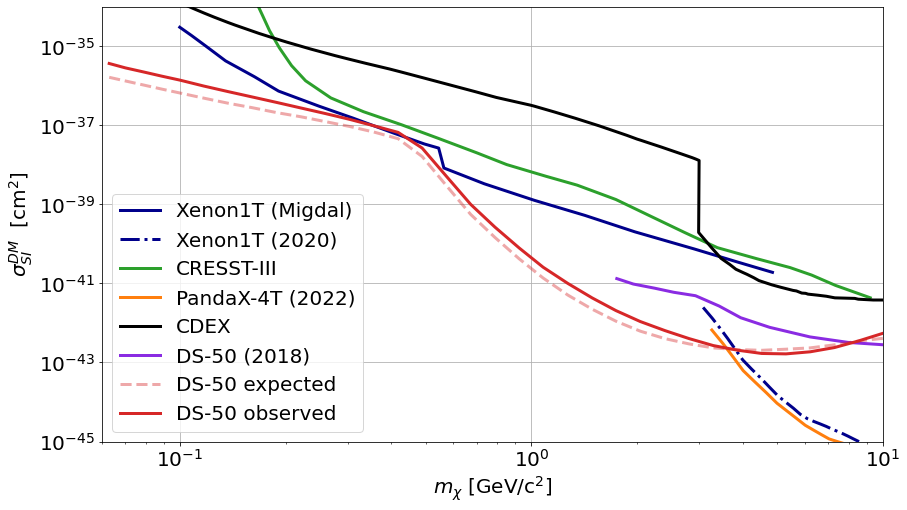}
    \caption{The $90\%\:{\rm C.I.}$ observed limit on the DM cross-section $\sigma_{SI}^{DM}$ as a function of the DM mass (red solid line) with the inclusion of the Migdal effect for the final \DSfs\ exposure of $12202\pm180$ kg d. The blue line is the frequentist 90\% C.L. from the XENON1T Migdal search \cite{Aprile:2019jmx}, the blue dashed-dot line is the one from the XENON1T NR search~\cite{Aprile:2019xxb}, the green line is the one by the CRESST-III experiment \cite{Abdelhameed:2019hmk}, the orange line is the one by the PandaX-4T experiment~\cite{PandaX-4T:2021bab}, the black line is the one by the CDEX experiment~\cite{CDEX:2018lau,CDEX:2021cll}, and the purple line is the one from the 2018 \DSfs\ pure NR analysis~\cite{Agnes:2018ves}. The pink dashed line is the expected sensitivity.}
    \label{fig:observed_sensitivity}
\end{figure}

Figure~\ref{fig:observed_sensitivity}
shows the $90\%\:{\rm C.I.}$ expected sensitivity on the DM cross-section $\sigma_{SI}^{DM}$ as a function of the DM mass (pink dashed line), using the Asimov (background only) dataset of Fig.~\ref{fig:pseudodata}.

The observed limit is shown as a red solid line in Fig.~\ref{fig:observed_sensitivity}. It is computed as the $90\%\:{\rm C.I.}$ bound on the DM cross-section $\sigma_{SI}^{DM}$ as a function of the DM mass using the full \DSfs\ dataset.

This line is compared with the most competitive exclusion limits, defined as frequentist 90\% C.L.s, in the GeV and sub-GeV DM mass region, i.e. the XENON1T Migdal search~\cite{Aprile:2019jmx} (blue line), the XENON1T NR search \cite{Aprile:2019xxb} (blue dashed-dot line), the CRESST-III result~\cite{Abdelhameed:2019hmk} (green line),
the PandaX-4T result~\cite{PandaX-4T:2021bab} (orange line), and the CDEX result~\cite{CDEX:2018lau,CDEX:2021cll} (black line).

No relevant deviations with respect to the expected sensitivity are observed. We performed the study of the behavior of the posterior \emph{pdf} of the nuisance parameters as a function of the mass of the DM candidate, obtaining similar results, reported in~\ref{app:exp_full_pos}.

\section{Conclusions}
\label{sec:conclusions}
In this work, we illustrated a novel technique, based on Bayesian Networks, to explicitly include the detector response model in the likelihood function of a measurement. 
This approach has the advantage with respect to the more conventional profile likelihood methods to not assume ``Gaussianity'' of the \emph{pdf}s nor linearity of the problem, to not rely on any signal or background spectrum morphing, and to keep the dependence of the parameter of interest on the response model parameters.
In this way, the result of the analysis is not only the measurement of the parameter of interest, but also a possible constraint on the response model. 
We deployed this method to the search for low mass dark matter with the \DSfs\ experiment and showed that  we obtain consistent results with the recently published analysis \cite{DarkSide-50:2022qzh}, and we further constrain the parameters of the detector response model.

In the first sections of this work we introduced the Bayesian Networks approach to the description of a multidimensional \emph{pdf}s. Such a method gives a clear picture of the connections between the different variables involved in the problem, allowing to identify groups of variable that decouple for the rest the problem. Figure~\ref{fig:Complete-Network} describes the entire likelihood of the measurement and shows the role of the various parameters and the information flow from one another.
The description of the problem in terms of BN allowed us to develop an inference algorithm based on a Markov Chain Monte Carlo (MCMC) to compute the posterior probability. This probability is constrained on the observed data, while retaining the dependence on each single parameter. 
A clever description of the detector response model in terms of parametric matrices allows the exploration of the impact of systematic variations of any parameter onto the final results. 
The proposed treatment significantly improves the understanding of the interplay between calibrations and spectra, as we elaborated in Sec.~\ref{sec:detector-effects} and as it is visible in Fig.~\ref{fig:bkg-spectra-syst}.
In addition, we provide a further insight on the systematic uncertainties induced by the detector model, and, using the additional constraining power of the analyzed data, reduce the uncertainty on several parameters of the detector model.

\begin{acknowledgements}
The DarkSide Collaboration offers its profound gratitude to the LNGS and its staff for their invaluable technical and logistical support. We also thank the Fermilab Particle Physics, Scientific, and Core Computing Divisions. Construction and operation of the DarkSide-50 detector was supported by the U.S. National Science Foundation (NSF) (Grants No. PHY-0919363, No. PHY-1004072, No. PHY-1004054, No. PHY-1242585, No. PHY-1314483, No. PHY-1314501, No. PHY-1314507, No. PHY-1352795, No. PHY-1622415, and associated collaborative grants No. PHY-1211308 and No. PHY-1455351), the Italian Istituto Nazionale di Fisica Nucleare, the U.S. Department of Energy (Contracts No. DE-FG02-91ER40671, No. DEAC02-07CH11359, and No. DE-AC05-76RL01830), the Polish NCN (Grant No. UMO-2019/33\-/B/ST2/02884) and the Polish Ministry for Education and Science (Grant No. 6811/IA/SP/2018). We also acknowledge financial support from the French Institut National de Physique Nucl\'eaire et de Physi\-que des Particules (IN2P3),   the  IN2P3-COPIN consortium (Grant No. 20-152),  and the UnivEarthS LabEx program (Grants No. ANR-10-LABX-0023 and No. ANR-18-IDEX-0001),  from the São Paulo Research Foundation (FAPESP) (Grant No. 2016/09084-0),  from the Interdisciplinary Scientific and Educational School of Moscow University ``Fundamental and Applied Space Research'',  from the Program of the Ministry of Education and Science of the  Russian  Federation  for  higher  education  establishments,  project No. FZWG-2020-0032 (2019-1569),  from IRAP AstroCeNT funded by FNP from ERDF, and from the Science and Technology Facilities Council, United Kingdom.  I.~Albuquerque is partially supported by the Brazilian Research Council (CNPq). This project has received funding from the European Union’s Horizon 2020 research and innovation program under grant agreement No 952480. Isotopes used in this research were supplied by the United States Department of Energy Office of Science by the Isotope Program in the Office of Nuclear Physics.

This version of the article has been accepted for publication, after peer review but is not the Version of Record and does not reflect post-acceptance improvements, or any corrections. The Version of Record is available online at\\
\href{http://dx.doi.org/10.1140/epjc/s10052-023-11410-4}{http://dx.doi.org/10.1140/epjc/s10052-023-11410-4}.
 \end{acknowledgements}

\begin{appendices}

\section{Expected joint posterior \emph{pdf}}
\label{app:exp_full_pos}

\begin{figure*}[!ht]
    \centering
    \includegraphics[width = \textwidth]{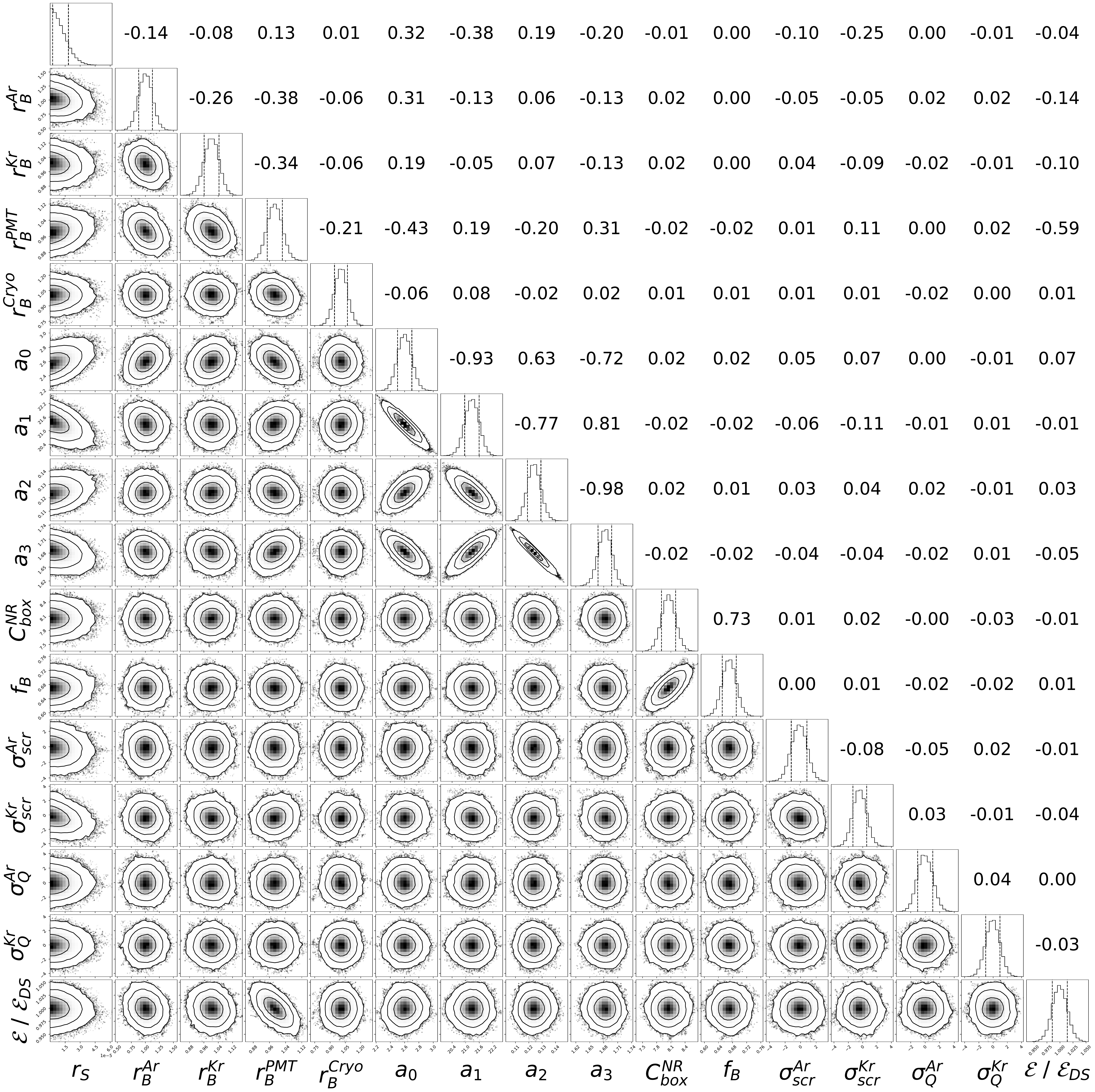}
    \caption{Graphical representation of the joint posterior \emph{pdf} of the fit on the expected pseudo-dataset assuming a DM mass $m_\chi = 4.5\:{\rm GeV}/c^2$. The plots on the diagonal of the figure are the uni-dimensional \emph{pdf} of each single parameter obtained by marginalizing on all the others. The bi-dimensional \emph{pdf}s in the bottom-left corner of the figure give the joint \emph{pdf}s of each pair of parameters obtained by marginalizing on the others. The plots show also the credible regions at $68\%,\:95\%,\:99.7\%$ probability as solid contour lines. The correlation coefficients are given in the upper-right corner of the figure.}
    \label{fig:4.52_expected_posterior}
\end{figure*}

\begin{figure*}
    \centering
    \includegraphics[width = 0.3\textwidth]{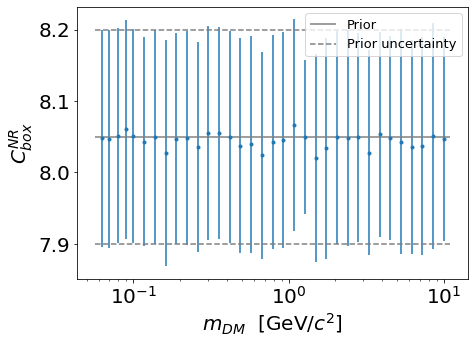}
    \includegraphics[width = 0.3\textwidth]{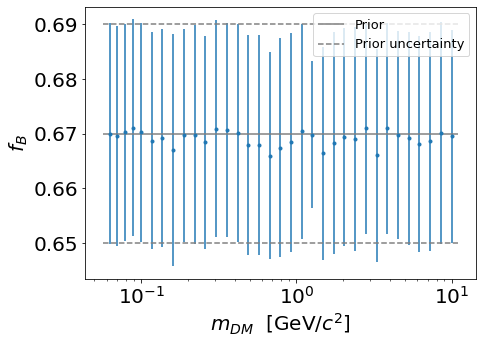}
    \includegraphics[width = 0.3\textwidth]{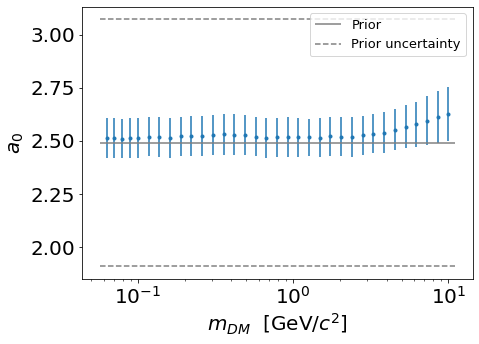}
    \includegraphics[width = 0.3\textwidth]{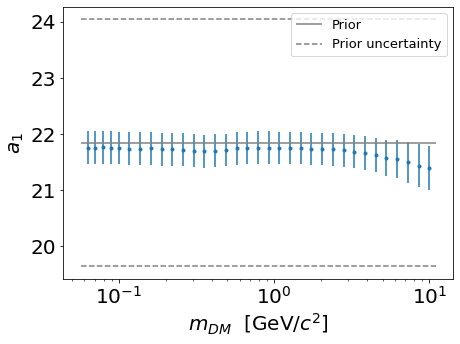}
    \includegraphics[width = 0.3\textwidth]{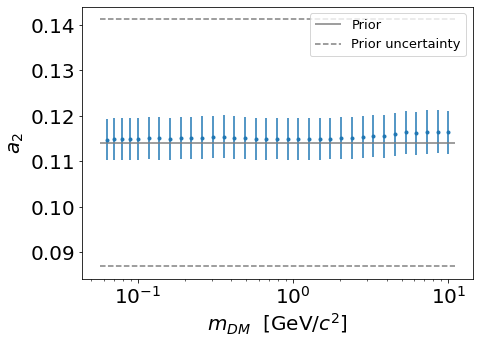}
    \includegraphics[width = 0.3\textwidth]{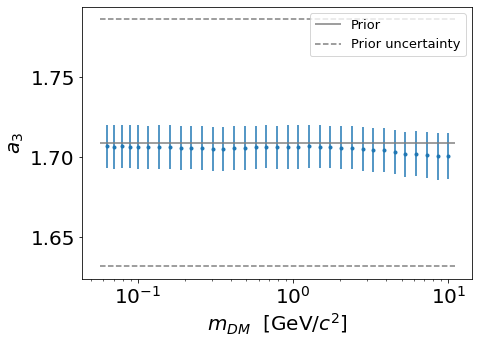}
    \includegraphics[width = 0.3\textwidth]{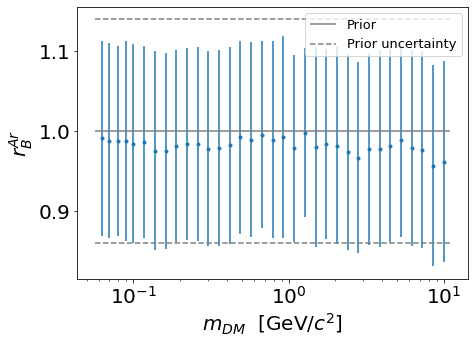}
    \includegraphics[width = 0.3\textwidth]{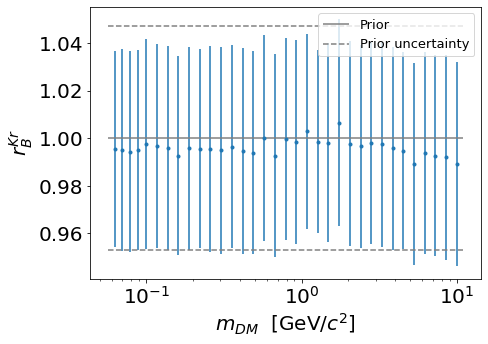}
    \includegraphics[width = 0.3\textwidth]{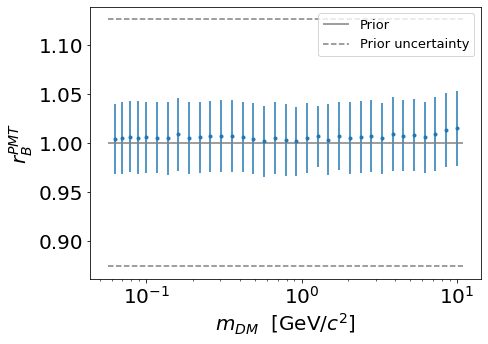}
    \includegraphics[width = 0.3\textwidth]{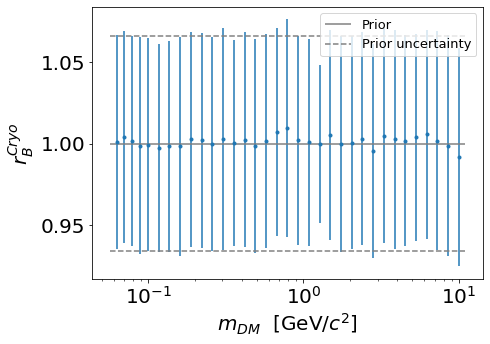}
    \includegraphics[width = 0.3\textwidth]{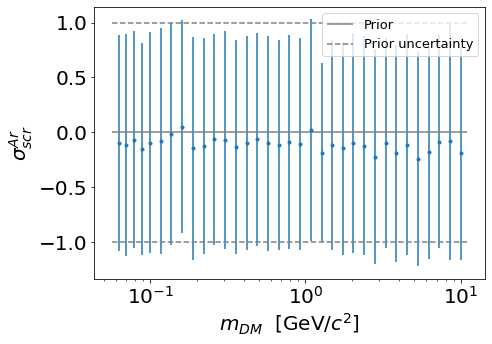}
    \includegraphics[width = 0.3\textwidth]{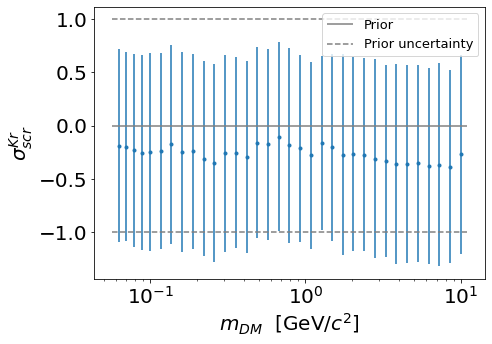}
    \includegraphics[width = 0.3\textwidth]{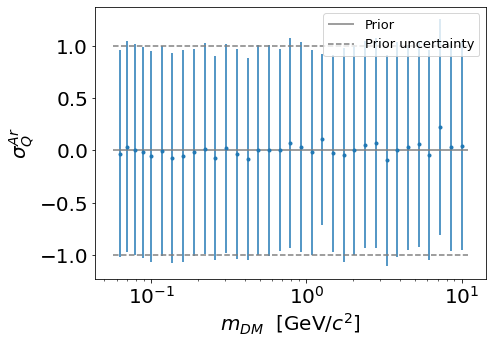}
    \includegraphics[width = 0.3\textwidth]{figs/Fig12/Q_ar.png}
    \includegraphics[width = 0.3\textwidth]{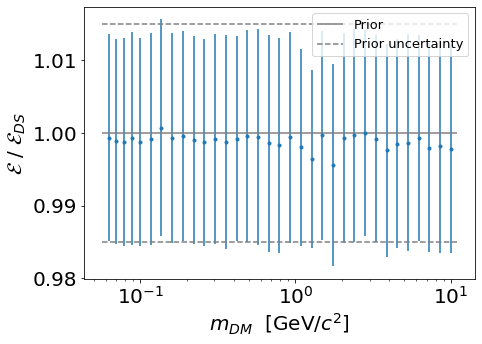}
    \caption{Evolution of the posterior \emph{pdf}s as a function of the DM mass. The blue points represent the mean of the posterior \emph{pdf} and their uncertainty bar is the correspondent standard deviation. In each plot, the gray solid line is the mean of the prior \emph{pdf}, while the two gray dashed lines correspond to the standard deviation of the prior \emph{pdf}.}
    \label{fig:systevolution_with_mass}
\end{figure*}

In Fig.~\ref{fig:4.52_expected_posterior} we report the joint posterior \emph{pdf} of the fit on the expected pseudo-dataset assuming a DM mass $m_\chi = 4.5\:{\rm GeV}/c^2$, and the evolution of the posterior \emph{pdf}s as a function of the mass is reported in Fig.~\ref{fig:systevolution_with_mass}.

% \clearpage
\section{Background only joint posterior \emph{pdf}}
\label{app:bkgonly_full_pos}

\begin{figure*}
    \centering
    \includegraphics[width = \textwidth]{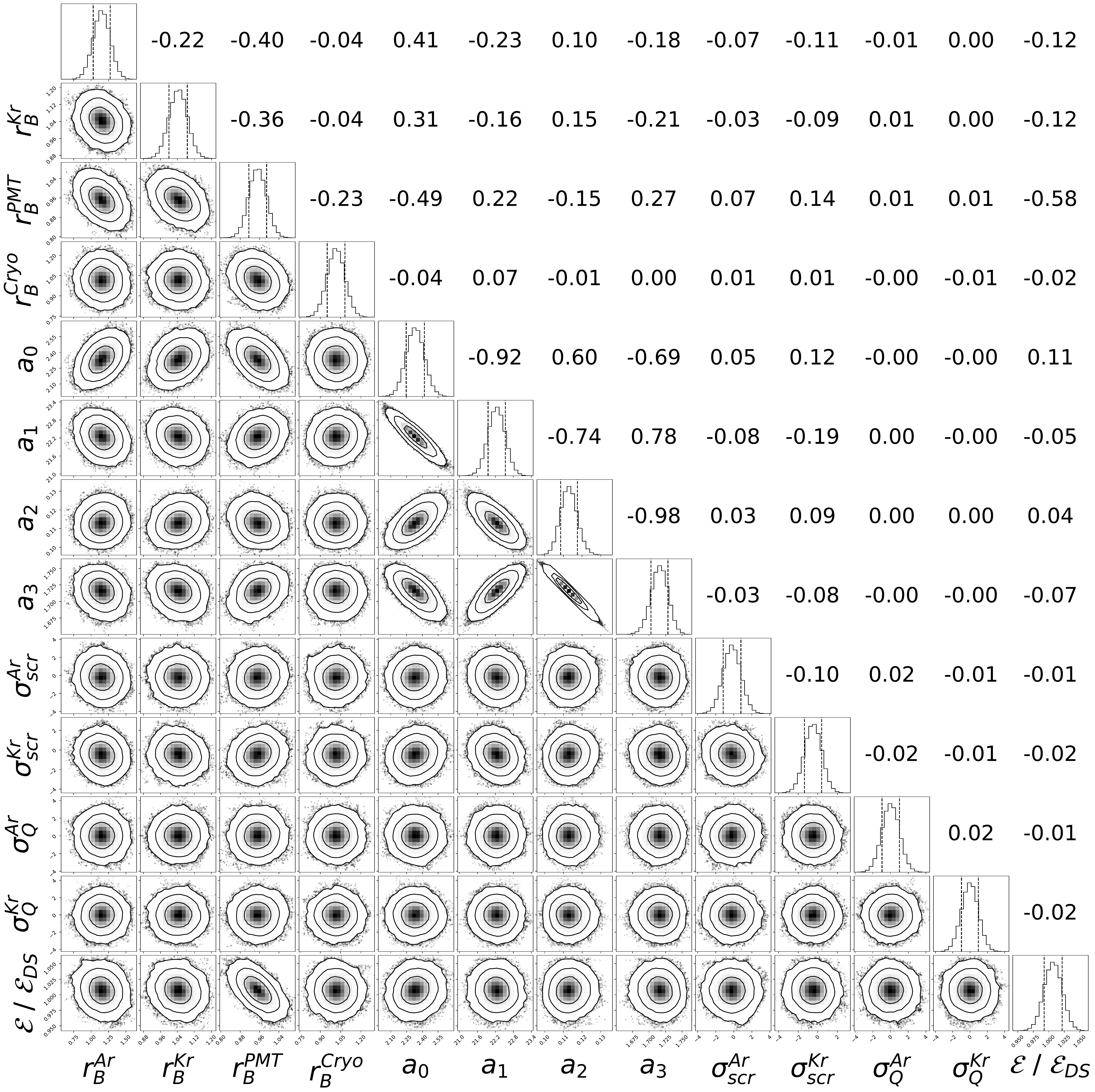}
    \caption{Graphical representation of the joint posterior \emph{pdf} of the background only fit on the \DSfs\ observed dataset. The plots on the diagonal of the figure are the uni-dimensional \emph{pdf} of each single parameter obtained by marginalizing on all the others. The bi-dimensional \emph{pdf}s in the bottom-left corner of the figure give the joint \emph{pdf}s of each pair of parameters obtained by marginalizing on the others. The plots show also the credible regions at $68\%,\:95\%,\:99.7\%$ probability as solid contour lines. The correlation coefficients are given in the upper-right corner of the figure.}
    \label{fig:bkgonly-posterior}
\end{figure*}

In Fig.~\ref{fig:bkgonly-posterior} we report joint posterior \emph{pdf} of the background only fit on the \DSfs\ observed dataset.

\end{appendices}

\newpage

\textcolor{white}{.}
\newpage
\clearpage

\bibliographystyle{jhep}       % APS-like style for physics
\bibliography{main}   % name your BibTeX data base

\newcommand{\notds}{\nolinebreak\footnotemark\nolinebreak}
\renewcommand{\thefootnote}{$*$}
{
\onecolumn
% \textbf{The Aria and DarkSide-20k Collaborations}
\textbf{The DarkSide-50 Collaboration}

P.~Agnes\thanksref{rhul}\nolinebreak,
I.F.M.~Albuquerque\thanksref{usp}\nolinebreak,
T.~Alexander\thanksref{pnnl}\nolinebreak,
A.K.~Alton\thanksref{augustana}\nolinebreak,
M.~Ave\thanksref{usp}\nolinebreak,
H.O.~Back\thanksref{pnnl}\nolinebreak,
G.~Batignani\thanksref{piinfn}{piuniphy}\nolinebreak,
K.~Biery\thanksref{fnal}\nolinebreak,
V.~Bocci\thanksref{rmunoinfn}\nolinebreak,
W.M.~Bonivento\thanksref{cainfn}\nolinebreak,
B.~Bottino\thanksref{geuni}{geinfn}\nolinebreak,
S.~Bussino\thanksref{rmtreinfn}{rmtreuni}\nolinebreak,
M.~Cadeddu\thanksref{cainfn}\nolinebreak,
M.~Cadoni\thanksref{cauniphy}{cainfn}\nolinebreak,
F.~Calaprice\thanksref{princeton}\nolinebreak,
A.~Caminata\thanksref{geinfn}\nolinebreak,
M.D.~Campos\thanksref{kings}\nolinebreak,
N.~Canci\thanksref{aqlngs}\nolinebreak,
M.~Caravati\thanksref{cainfn}\nolinebreak,
N. Cargioli\thanksref{cainfn}\nolinebreak,
M.~Cariello\thanksref{geinfn}\nolinebreak,
M.~Carlini\thanksref{aqlngs}{aqgssi}\nolinebreak,
V.~Cataudella\thanksref{nauniphy}{nainfn}\nolinebreak,
P.~Cavalcante\thanksref{vtech}{aqlngs}\nolinebreak,
S.~Cavuoti\thanksref{nauniphy}{nainfn}\nolinebreak,
S.~Chashin\thanksref{msu}\nolinebreak,
A.~Chepurnov\thanksref{msu}\nolinebreak,
C.~Cical\`o\thanksref{cainfn}\nolinebreak,
G.~Covone\thanksref{nauniphy}{nainfn}\nolinebreak,
D.~D'Angelo\thanksref{miuni}{miinfn}\nolinebreak,
S.~Davini\thanksref{geinfn}\nolinebreak,
A.~De~Candia\thanksref{nauniphy}{nainfn}\nolinebreak,
S.~De~Cecco\thanksref{rmunoinfn}{rmunouni}\nolinebreak,
G.~De~Filippis\thanksref{nauniphy}{nainfn}\nolinebreak,
G.~De~Rosa\thanksref{nauniphy}{nainfn}\nolinebreak,
A.V.~Derbin\thanksref{petersburg}\nolinebreak,
A.~Devoto\thanksref{cauniphy}{cainfn}\nolinebreak,
M.~D'Incecco\thanksref{aqlngs}\nolinebreak,
C.~Dionisi\thanksref{rmunoinfn}{rmunouni}\nolinebreak,
F.~Dordei\thanksref{cainfn}\nolinebreak,
M.~Downing\thanksref{umass}\nolinebreak,
D.~D'Urso\thanksref{ssunichp}{ctlns}\nolinebreak,
M.~Fairbairn\thanksref{kings}\nolinebreak,
G.~Fiorillo\thanksref{nauniphy}{nainfn}\nolinebreak,
D.~Franco\thanksref{apc}\nolinebreak,
F.~Gabriele\thanksref{cainfn}\nolinebreak,
C.~Galbiati\thanksref{princeton}{aqgssi}{aqlngs}\nolinebreak,
C.~Ghiano\thanksref{aqlngs}\nolinebreak,
C.~Giganti\thanksref{lpnhe}\nolinebreak,
G.K.~Giovanetti\thanksref{princeton}\nolinebreak,
A.M.~Goretti\thanksref{aqlngs}\nolinebreak,
G.~Grilli di Cortona\thanksref{lnfinfn}{rmunoinfn}\nolinebreak,
A.~Grobov\thanksref{kurchatov}{mephi}\nolinebreak,
M.~Gromov\thanksref{msu}{jinr}\nolinebreak,
M.~Guan\thanksref{ihep}\nolinebreak,
M.~Gulino\thanksref{enunicee}{ctlns}\nolinebreak,
B.R.~Hackett\thanksref{pnnl}\nolinebreak,
K.~Herner\thanksref{fnal}\nolinebreak,
T.~Hessel\thanksref{apc}\nolinebreak,
B.~Hosseini\thanksref{cainfn}\nolinebreak,
F.~Hubaut\thanksref{cppm}\nolinebreak,
E.V.~Hungerford\thanksref{houston}\nolinebreak,
An.~Ianni\thanksref{princeton}{aqlngs}\nolinebreak,
V.~Ippolito\thanksref{rmunoinfn}\nolinebreak,
K.~Keeter\thanksref{bhsu}\nolinebreak,
C.L.~Kendziora\thanksref{fnal}\nolinebreak,
M.~Kimura\thanksref{astrocent}\nolinebreak,
I.~Kochanek\thanksref{aqlngs}\nolinebreak,
D.~Korablev\thanksref{jinr}\nolinebreak,
G.~Korga\thanksref{houston}{aqlngs}\nolinebreak,
A.~Kubankin\thanksref{belgorod}\nolinebreak,
M.~Kuss\thanksref{piinfn}\nolinebreak,
M.~La~Commara\thanksref{nauniphy}{nainfn}\nolinebreak,
M.~Lai\thanksref{cauniphy}{cainfn}\nolinebreak,
X.~Li\thanksref{princeton}\nolinebreak,
M.~Lissia\thanksref{cainfn}\nolinebreak,
G.~Longo\thanksref{nauniphy}{nainfn}\nolinebreak,
O.~Lychagina\thanksref{jinr}{msu}\nolinebreak,
I.N.~Machulin\thanksref{kurchatov}{mephi}\nolinebreak,
L.P.~Mapelli\thanksref{ucla}\nolinebreak,
S.M.~Mari\thanksref{rmtreinfn}{rmtreuni}\nolinebreak,
J.~Maricic\thanksref{hawaii}\nolinebreak,
A.~Messina\thanksref{rmunoinfn}{rmunouni}\nolinebreak,
R.~Milincic\thanksref{hawaii}\nolinebreak,
J.~Monroe\thanksref{rhul}\nolinebreak,
M.~Morrocchi\thanksref{piinfn}{piuniphy}\nolinebreak,
X.~Mougeot\thanksref{lnhb}\nolinebreak,
V.N.~Muratova\thanksref{petersburg}\nolinebreak,
P.~Musico\thanksref{geinfn}\nolinebreak,
A.O.~Nozdrina\thanksref{kurchatov}{mephi}\nolinebreak,
A.~Oleinik\thanksref{belgorod}\nolinebreak,
F.~Ortica\thanksref{pgunicbb}{pginfn}\nolinebreak,
L.~Pagani\thanksref{ucdavis}\nolinebreak,
M.~Pallavicini\thanksref{geuni}{geinfn}\nolinebreak,
L.~Pandola\thanksref{ctlns}\nolinebreak,
E.~Pantic\thanksref{ucdavis}\nolinebreak,
E.~Paoloni\thanksref{piinfn}{piuniphy}\nolinebreak,
K.~Pelczar\thanksref{aqlngs}{krakow}\nolinebreak,
N.~Pelliccia\thanksref{pgunicbb}{pginfn}\nolinebreak,
S.~Piacentini\thanksref{rmunoinfn}{rmunouni}\nolinebreak,
A.~Pocar\thanksref{umass}\nolinebreak,
D.M.~Poehlmann\thanksref{ucdavis}\nolinebreak,
S.~Pordes\thanksref{fnal}\nolinebreak,
S.S.~Poudel\thanksref{houston}\nolinebreak,
P.~Pralavorio\thanksref{cppm}\nolinebreak,
D.D.~Price\thanksref{manchester}\nolinebreak,
F.~Ragusa\thanksref{miuni}{miinfn}\nolinebreak,
M.~Razeti\thanksref{cainfn}\nolinebreak,
A.~Razeto\thanksref{aqlngs}\nolinebreak,
A.L.~Renshaw\thanksref{houston}\nolinebreak,
M.~Rescigno\thanksref{rmunoinfn}\nolinebreak,
J.~Rode\thanksref{lpnhe}{apc}\nolinebreak,
A.~Romani\thanksref{pgunicbb}{pginfn}\nolinebreak,
D.~Sablone\thanksref{princeton}{aqlngs}\nolinebreak,
O.~Samoylov\thanksref{jinr}\nolinebreak,
E.~Sandford\thanksref{manchester}\nolinebreak,
W.~Sands\thanksref{princeton}\nolinebreak,
S.~Sanfilippo\thanksref{ctlns}\nolinebreak,
C.~Savarese\thanksref{princeton}\nolinebreak,
B.~Schlitzer\thanksref{ucdavis}\nolinebreak,
D.A.~Semenov\thanksref{petersburg}\nolinebreak,
A.~Shchagin\thanksref{belgorod}\nolinebreak,
A.~Sheshukov\thanksref{jinr}\nolinebreak,
M.D.~Skorokhvatov\thanksref{kurchatov}{mephi}\nolinebreak,
O.~Smirnov\thanksref{jinr}\nolinebreak,
A.~Sotnikov\thanksref{jinr}\nolinebreak,
S.~Stracka\thanksref{piinfn}\nolinebreak,
Y.~Suvorov\thanksref{nauniphy}{nainfn}\nolinebreak,
R.~Tartaglia\thanksref{aqlngs}\nolinebreak,
G.~Testera\thanksref{geinfn}\nolinebreak,
A.~Tonazzo\thanksref{apc}\nolinebreak,
E.V.~Unzhakov\thanksref{petersburg}\nolinebreak,
A.~Vishneva\thanksref{jinr}\nolinebreak,
R.B.~Vogelaar\thanksref{vtech}\nolinebreak,
M.~Wada\thanksref{astrocent}{cauniphy}\nolinebreak,
H.~Wang\thanksref{ucla}\nolinebreak,
Y.~Wang\thanksref{ucla}{ihep}\nolinebreak,
S.~Westerdale\thanksref{ucr}\nolinebreak,
M.M.~Wojcik\thanksref{krakow}\nolinebreak,
X.~Xiao\thanksref{ucla}\nolinebreak,
C.~Yang\thanksref{ihep}\nolinebreak,
G.~Zuzel\thanksref{krakow}

\begin{enumerate}[label=\textsuperscript{\arabic*}]
    \item {\label{rhul}\RHUL}
    \item {\USP\label{usp}}
    \item {\PNNL\label{pnnl}}
    \item {\Augustana\label{augustana}}
    \item {\PIINFN\label{piinfn}}
    \item {\PIUniPHY\label{piuniphy}}
    \item {\FNAL\label{fnal}}
    \item {\RMUnoINFN\label{rmunoinfn}}
    \item {\CAINFN\label{cainfn}}
    \item {\GEUni\label{geuni}}
    \item {\GEINFN\label{geinfn}}
    \item {\RMTreINFN\label{rmtreinfn}}
    \item {\RMTreUni\label{rmtreuni}}
    \item {\CAUniPHY\label{cauniphy}}
    \item {\Princeton\label{princeton}}
    \item {\kings\label{kings}}
    \item {\AQLNGS\label{aqlngs}}
    \item {\AQGSSI\label{aqgssi}}
    \item {\SSUniCHP\label{ssunichp}}
    \item {\CTLNS\label{ctlns}}
    \item {\NAUniPHY\label{nauniphy}}
    \item {\NAINFN\label{nainfn}}
    \item {\VTech\label{vtech}}
    \item {\MSU\label{msu}}
    \item {\MIUni\label{miuni}}
    \item {\MIINFN\label{miinfn}}
    \item {\RMUnoUni\label{rmunouni}}
    \item {\Petersburg\label{petersburg}}
    \item {\UMass\label{umass}}
    \item {\APC\label{apc}}
    \item {\LPNHE\label{lpnhe}}
    \item {\LNFINFN\label{lnfinfn}}
    \item {\Kurchatov\label{kurchatov}}
    \item {\MEPhI\label{mephi}}
    \item {\JINR\label{jinr}}
    \item {\IHEP\label{ihep}}
    \item {\ENUniCEE\label{enunicee}}
    \item {\CPPM\label{cppm}}
    \item {\Houston\label{houston}}
    \item {\BHSU\label{bhsu}}
    \item {\AstroCeNT\label{astrocent}}
    \item {\Belgorod\label{belgorod}}
    \item {\UCLA\label{ucla}}
    \item {\Hawaii\label{hawaii}}
    \item {\LNHB\label{lnhb}}
    \item {\PGUniCBB\label{pgunicbb}}
    \item {\PGINFN\label{pginfn}}
    \item {\UCDavis\label{ucdavis}}
    \item {\Krakow\label{krakow}}
    \item {\Manchester\label{manchester}}
    \item {\UCRiverside\label{ucr}}
\end{enumerate}

}

\end{document}